\documentclass[superscriptaddress,groupedaddress,nofootnoteinbib,11pt]{article}  % for review and submission

\usepackage{graphicx}  % needed for figures
\usepackage{dcolumn}   % needed for some tables
\usepackage{bm}        % for math
\usepackage{slashed}        % for math
\usepackage{amsmath}
\usepackage{mathtools}
\usepackage{jcappub}

\def\be{\begin{equation}}
\def\ee{\end{equation}}
\def\ba{\begin{eqnarray}}
\def\ea{\end{eqnarray}}

\def\e{{\epsilon}}

\def\L*{{\cal L}_*}
\def\L{\mathcal{L}}
\def\R{\mathcal{R}}
\def\P{\mathcal{P}}
\def\({\left(}
\def\){\right)}

\def\<{\langle}
\def\>{\rangle}

\def\cs2{c_{s}^{2}}

\begin{document}

\title{Correlating CMB Spectral Distortions with Temperature: what do we learn on Inflation?}

\author{Emanuela Dimastrogiovanni$^a$ and} \author{Razieh Emami$^b$}

\affiliation{$^a$Department of Physics and School of Earth and Space Exploration, Arizona State University, Tempe, AZ 85827}
\affiliation{$^b$  Institute for Advanced Study, The Hong Kong University of Science and Technology, Clear Water Bay, Kowloon, Hong Kong}

\date{\today}

\abstract{Probing correlations among short and long-wavelength cosmological fluctuations is known to be decisive for deepening the current understanding of inflation at the microphysical level. Spectral distortions of the CMB can be caused by dissipation of cosmological perturbations when they re-enter Hubble after inflation. Correlating spectral distortions with temperature anisotropies will thus provide the opportunity to greatly enlarge the range of scales over which squeezed limits can be tested, opening up a new window on inflation complementing the ones currently probed with CMB and LSS.   \\
In this paper we discuss a variety of inflationary mechanisms that can be efficiently constrained with distortion-temperature correlations. For some of these realizations (representative of large classes of models) we derive quantitative predictions for the squeezed limit bispectra, finding that their amplitudes are above the sensitivity limits of an experiment such as the proposed PIXIE.
}

\maketitle

%sottolinea mu-y enhancement nell'introduzione, anche per sottolineare che la scelta che facciamo dei due modelli e' dettata dal voler presentare due esempi diversi (intermediate vs monotonic)

\section{Introduction}

Inflation \cite{inflation} offers a physical picture for the origin of cosmic structures that is firmly supported by observations. While we have successfully tested its paradigm \cite{tested}, the microphysics of inflation is largely unknown. New cosmological surveys have been purposely designed to address this fundamental gap in our understanding of early Universe physics. A crucial observable that is related to the particle content of the primordial Universe is non-Gaussianity. This refers to the information contained in correlation functions beyond the power spectrum. The tightest bounds to date on the non-Gaussianity of primordial density fluctuations are provided by large-angle Cosmic Microwave Background (CMB) anisotropies. These have been able to constrain the bispectrum amplitude in various momentum configurations with a sensitivity of $\Delta f_{\text{nl}}\gtrsim\mathcal{O}(10)$ \cite{Ade:2015ava}. Upcoming galaxy surveys aim at $\Delta f_{\text{nl}}\simeq\mathcal{O}(1)$ \cite{rev}, a significant theoretical thresholds that marks the separation between fundamentally different mechanisms for inflation. \\

Non-Gaussianity from squeezed momenta configurations, i.e. arising from coupling among modes of widely different wavelengths, is particularly important for model-building. Its detection would rule out all inflationary models where primordial fluctuations are entirely determined by the time-evolution of a single degree of freedom (single-clock), favoring a more complex dynamics. Squeezed non-Gaussianity thus represents a unique probe for the spectrum of masses and interactions during inflation. Another critical implication of a physical coupling in the squeezed limit would be the presence of a super-cosmic variance bias for local observables \cite{LoVerde:2013xka}. The latter would need to be accounted for in order to correctly match observations to the inflationary Lagrangian.\\  

A number of directions are being developed for testing squeezed non-Gaussianity in the large scale structure (LSS). Besides the bispectrum, observables that are highly sensitive to the existence of such a coupling include the scale-dependent halo bias \cite{Dalal:2007cu}. Fossil-type signatures, in the form of an off-diagonal correlation between different Fourier modes of the density field \cite{fossils}, are also highly sensitive to the presence of a long-short wavelength coupling.  \\
\indent Squeezed non-Gaussianity can be probed on a vast range of scales: searches on large-angle CMB scales and in the LSS ($10^{-4}\, \text{Mpc}^{-1}\lesssim k\lesssim 1\, \text{Mpc}^{-1}$) will be complemented by CMB spectral distortions observations. The latter provide access to scales up to $k\sim 10^{4}\, \text{Mpc}^{-1}$, opening up a window over 10 additional inflationary e-folds \cite{damping}. The black-body spectrum of the CMB can be distorted by any mechanism resulting in an exchange of energy (or photons) with the CMB after $z\simeq 2\times 10^6$ (at earlier times, thermalization processes would be highly efficient and any produced distortion would be quickly erased) \cite{rev2}. One may use spectral distortion to shed light on both the early and late universe, investigating processes such as cosmological recombination \cite{rec}, reionization and structure formation \cite{reion-stf}, decaying or annihilating particles \cite{dec}, initial states for inflation \cite{istates}, primordial anisotropies \cite{Shiraishi:2015lma}, primordial black holes \cite{pbh}, and other possibilities. Spectral distortion also arises from dissipation of primordial fluctuations. Super-horizon fluctuations crossing inside the Hubble radius after inflation are damped by photon diffusion when their wavelengths fall below the photon mean free path. Specifically, modes with wave-numbers between $1$ and $50\,\text{Mpc}^{-1}$ dissipate during the $y$ distortion era ($z\lesssim 5\times 10^4$), modes between $50$ and $10^{4}\,\text{Mpc}^{-1}$ dissipate during the $\mu$ distortion era ($2\times 10^6\gtrsim z \gtrsim 5\times 10^4$). The power spectrum of primordial fluctuations on these scales can thus be probed through spectral distortion \cite{powers}. A correlation of the anisotropic distortion with temperature anisotropies is a direct probe of a squeezed primordial bispectrum as pointed out for the first time in \cite{bisp} for $\mu$ distortion, whereas distortion-distortion correlations probe primordial trispectra \cite{trisp}. \\
\indent In recent work, \cite{us1} (see also \cite{us2}), we stress the importance of also including $y$ distortion in this context, not only to enlarge the range of scales over which non-Gaussianity is probed (which would eventually allow to test its scale-dependence without any a-priori parameterization), but to also improve on the chances of detection if $f_{\text{nl}}$ were larger on y scales compared to $\mu$ or large-angle CMB scales. Here a forecast for $\Delta f_{\text{nl}}$ was provided on $\mu$ and on $y$ distortion scales for a bispectrum template common to a number of scenarios. \\
\indent An interesting open problem on the data-analysis front is envisaging techniques \cite{Zhang:2004fh} that would allow to disentangle the $y$ signal of primordial fluctuations from the one produced at late times by the Sunyaev-Zel'dovich effect. Another important question one should address in order to understand the extent to which near-future experiments targeting spectral distortion can test inflation through temperature-distortion correlations is to identify the general conditions that lead to observable levels of non-Gaussianity on $\mu$ or $y$ scales. We do so in this paper, deriving also quantitative predictions for some interesting realizations that are representative of wide classes of models.  \\

This paper unfolds as follows: in Secs.~\ref{general1} and \ref{general2} we briefly review how to probe inflationary observables on small-scales with spectral distortions and discuss some inflationary mechanisms naturally predicting a squeezed bispectrum that is potentially detectable through temperature-distortion correlations; in Secs.~\ref{resonance} and \ref{g-hybrid} we analyze the predictions for scale-dependent non-Gaussianity (henceforth \textsl{sdnG}) from concrete realizations for, respectively, resonance production of cosmological fluctuations during inflation and the \textsl{gauged hybrid} mechanism (the former is representative of models where the $y-T$ correlation is larger than $\mu-T$, in the latter case instead $f_{\text{nl}}$ grows monotonically on small scales, leading to a distortion-temperature correlation that is largest on $\mu$ scales); we offer our conclusions in Sec.~\ref{conclusions}. More details on the derivation of three-point correlations in the resonance model are provided in Appendix A.

\section{Spectral distortion from diffusion damping: Gaussian and non-Gaussian observables}
\label{general1}

After inflation, the Hubble radius grows over time faster than physical scales and fluctuations that were initially stretched outside Hubble gradually re-enter. CMB photons undergo multiple scatterings in the primordial plasma, performing a random walk with a characteristic length scale, $\lambda_{D}(z)$. When the wavelength of fluctuations falls below the photons mean-free path, they are damped by photon diffusion. In the $\mu$ and $y$ era, the scales that undergo damping correspond, respectively, to modes $50\,\text{Mpc}^{-1}\lesssim k\lesssim 10^{4}\,\text{Mpc}^{-1}$ and to $1\,\text{Mpc}^{-1}\lesssim k\lesssim 50\,\text{Mpc}^{-1}$ \footnote{Distortions generated below $1\,\text{Mpc}^{-1}$, i.e. after recombination, are much smaller than those generated on smaller scales \cite{smaller}. Those produced on scales between $10^{4}$ and $10^5\,\text{Mpc}^{-1}$, on the other hand, generate entropy in the primordial plasma \cite{larger}.}. The damping of primordial fluctuations can be described as a mixing of photons with blackbody distributions of different temperatures (``hot'' and ``cold'' photons). The immediate result of a such a mixing is a $y$-distorted blackbody spectrum. This converts into a $\mu$-distorted one if Compton scattering is highly efficient. The resulting distortions are given by \cite{Chluba:2011hw}
\begin{eqnarray}
\langle \mu\rangle\approx\int d\log k\,\mathcal{P}_{\zeta}(k)\,W_{\mu}(k)\,,\\
\langle y\rangle\approx\int d\log k\,\mathcal{P}_{\zeta}(k)\,W_{y}(k)\,,
\end{eqnarray}
where $\mathcal{P}_{\zeta}(k)$ is the primordial scalar power spectrum and $W_{\mu}(k)$ and $W_{y}(k)$ are window functions encoding acoustic damping and thermalization effects. In the pre-recombination era, if one neglects the contribution of intermediate distortions (and assuming a sharp transition from $\mu$ to $y$ era), these can be well-approximated by 
\begin{eqnarray}
&&W_{\mu}(k)\approx 2.3\left[e^{-2k^2/k_{D}^{2}(z_{\mu})}-e^{-2k^2/k_{D}^{2}(z_{\mu y})}\right] \,,\\
&&W_{y}(k)\approx 0.4\left[e^{-2k^2/k_{D}^{2}(z_{\mu y})}-e^{-2k^2/k_{D}^{2}(z_{y})}\right] \,,
\end{eqnarray}
where $k_{D}\approx 4\times 10^{-6} (1+z)^{3/2}\,\text{Mpc}^{-1}$ is the diffusion damping scale, $z_{\mu}\approx 2 \times 10^6$ and $z_{\mu y}\approx 5\times 10^4$ mark, respectively, the beginning and end of the $\mu$ era, and $z_{y}\approx 1090$. \\

The sensitivity limits of an experiment like PIXIE are very close to the detection limit needed for a small-scale power spectrum with the same scale dependence as the one currently detected from CMB temperature anisotropies. As a result, a small deviation from the latter may be easily detected if it predicts additional power on small scales. This is the case in a large number of scenarios including, for instance, particle production, phase transitions, massive fields \cite{powers}.     \\

\begin{figure}
	\begin{center}
	\includegraphics[width=0.8\textwidth]{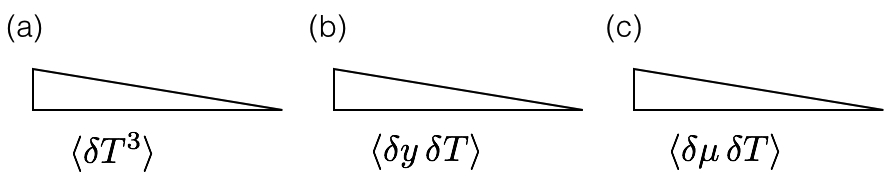}
	\caption{Representation of the bispectrum in the squeezed limit for: (a) temperature correlation (both long and short wavelength modes are on CMB scales); (b) y distortion-temperature correlation (short-wavelength modes are in the $[1,50]\,\text{Mpc}^{-1}$ range); (c) $\mu$ distortion-temperature correlation (short-wavelength modes are in the $[50,10^4]\,\text{Mpc}^{-1}$ range).}
		\label{FIG1}
	\end{center}
	\end{figure}

The fractional variation of the distortion is proportional to the convolution of two density perturbations. As a result, its correlations with the fluctuations in temperature corresponds to a primordial bispectrum in the squeezed limit. One can easily derive this result in the local model, where the curvature fluctuation is expanded as a power series of a Gaussian random variable as $\zeta(\vec{x})\approx z(\vec{x})+(3/5)f_{\text{nl}}\,z^{2}(\vec{x})$. Performing a long-short mode decomposition $\zeta=\zeta_{\text{L}}+\zeta_{\text{S}}$, one finds
\begin{equation}
\zeta_{\text{S}}(\vec{x})\approx z_{\text{S}}(\vec{x})\left[1+\frac{6}{5}\,f_{\text{nl}}\,\zeta_{\text{L}}(\vec{x})\right]\,.
\end{equation}
The fractional variation of the distortion ($d\equiv\mu,y$) is given by $\Delta d/d\approx \delta\langle \zeta_{\text{S}}^2\rangle/\langle\zeta_{\text{S}}^2\rangle\approx (12/5) \,f_{\text{nl}}\,\zeta_{\text{L}}$. Correlating the latter with the large-angle temperature fluctuation, $\Delta T/T\approx \zeta_{\text{L}}/5$ gives the following result in terms of the temperature autocorrelation
\begin{equation}
C_{l}^{d-T}\approx 12\,f_{\text{nl}}\, C_{l}^{TT}\,.
\end{equation}
For a squeezed-limit bispectrum template $\mathcal{B}_{\zeta}(k_{1},k_{2},k_{3})\approx (12/5)\,f_{\text{nl}}\,\mathcal{P}_{\zeta}(k_{\text{L}})\mathcal{P}_{\zeta}(k_{\text{S}})$, and under a few simplifying assumptions, one finds the following forecast for the smallest detectable amplitude on $\mu$ and $y$ scales \cite{us1}\footnote{Note that very recent findings \cite{Creque-Sarbinowski:2016wue} estimate that the minimum $f_{\text{nl}}$ amplitude on y-scales would be about two orders of magnitude larger if one accounts for the $y-T$ correlation generated at late times.}
\begin{eqnarray}
&&f_{\text{nl}}^{(\mu)}\simeq 10^2\left(\frac{\mu_{\text{min}}}{10^{-9}}\right)\left(\frac{\langle \mu\rangle}{2\times 10^{-8}}\right)^{-1}\,,\\
&&f_{\text{nl}}^{(y)}\simeq 10^2\left(\frac{y_{\text{min}}}{2\times10^{-10}}\right)\left(\frac{\langle y\rangle}{4\times 10^{-9}}\right)^{-1}\,,
\end{eqnarray}
where $\langle \mu\rangle$ and $\langle y\rangle$ are the smallest detectable monopoles for a given experiment. Even though they were derived for a sample bispectrum template, these preliminary forecasts suggest that an experiment with a sensitivity like PIXIE will only be able to detect non-Gaussianity on small scales if $f_{\text{nl}}$ on those scales is larger than the current large-scale upper bounds.  For this reason, in the rest of the paper we will be focusing on models predicting a squeezed-limit amplitude $f_{\text{nl}}$ that is significantly enhanced on small scales (as in configuration \textsl{b} or also \textsl{c} of Fig.~\ref{FIG1}) w.r.t. large scales (e.g. in configuration \textsl{a}).   \\

\section{Enhanced squeezed non-Gaussianity on small scales from inflation}
\label{general2}

For a primordial non-Gaussian signal on $y$ or $\mu$ scales to be possibly observable with upcoming experiments (e.g. PIXIE) two conditions should be met: (i) modes of widely different wavelengths should be correlated; (ii) the correlation should be enhanced on smaller scales w.r.t. typical CMB scales. In what follows we discuss a number of mechanisms satisfying these two conditions.\\

\subsection{Realizing a coupling between long and short wavelengths}
\label{sec31}

In standard single-field slow-roll inflation (SFRS), as in all single-clock inflation set-ups, the effect of a long-wavelength mode on short-wavelength modes is a (unobservable) background rescaling on the short modes \cite{Maldacena:2002vr}. This can be easily understood as follows. Consider for example the scalar bispectrum in the squeezed limit ($k_{3}\ll k_{1}\simeq k_{2}$). The $k_{3}$ mode crosses outside Hubble earlier than $k_{1}$ and $k_{2}$ and it is frozen by the time the other two modes reach horizon size (a correlation cannot arise earlier than that time in the basic SFSR scenario). As such, $k_{3}$ generates a rescaling of the background $a\rightarrow a\,e^{\zeta_{3}}\simeq a(1+\zeta_{3})$, which entails anticipating the horizon exit of $k_{1}$ and $k_{2}$ by an amount $\delta t\simeq -\zeta_{3}/H$. As a result, the two point correlation for the hard modes in the presence of the soft mode is $\langle \zeta_{1}\zeta_{2}\rangle_{\zeta_{3}}\sim(d\langle \zeta_{1}\zeta_{2}\rangle/dt)\delta t\sim(1/H)(d\langle \zeta_{1}\zeta_{2}\rangle/dt)\zeta_{3}$, from which it follows
\begin{equation}\label{eqq1}
\langle \zeta_{1}\zeta_{2}\zeta_{3}\rangle\sim(1-n_{s})P_{\zeta}(k_{1})P_{\zeta}(k_{3}) \,.
\end{equation}
This is the formulation, to leading-order, of the single-clock \textsl{consistency conditions} (ccs) for the scalar bispectrum, which can be also rewritten as 
\begin{equation}
B_{\zeta}(k_{L},k_{S},k_{S})\simeq -P_{\zeta}(k_{L})P_{\zeta}(k_{S})\frac{\partial\ln [k_{S}^{3}P_{\zeta}(k_{S})]}{\partial\ln k_{S}}\,,
\end{equation}
where $k_{3}\equiv k_{L}\ll k_{1}\simeq k_{2}\equiv k_{S}$. The result in Eq.~(\ref{eqq1}) follows from a gauge choice \cite{Dai:2013kra}. Projecting to the coordinate frame that is appropriate for late time observers, one finds 
\begin{equation}
B^{\text{obs}}(k_{L},k_{S},k_{S})\simeq B_{\zeta}(k_{L},k_{S},k_{S})+P_{\zeta}(k_{L})P_{\zeta}(k_{S})\frac{\partial\ln\left[k_{S}^{3}P_{\zeta}(k_{S})\right]}{\partial\ln k_{S}}+\mathcal{O}\left(\frac{k_{L}}{k_{S}}\right)^{2}\,.
\end{equation}
The r.h.s. of the previous equation is therefore equal to zero in single-clock inflation to leading order in $k_{L}/k_{S}$. For a physical (and potentially observable) primordial correlation in the squeezed limit to be generated, at least one of these two conditions (that are caveats in the derivation of Eq.~(\ref{eqq1})) characterizing single-clock inflation should be violated: (a) correlations cannot be generated on sub-horizon scales; (b) modes are frozen on super-horizon scales.  \\

An entirely reasonable expectation is that multiple fields were around during inflation and might have played a role in the background dynamics or by contributing to cosmological fluctuations. This is certainly the case, for instance, in the case of a string-theory embedding. In multi-field inflation, condition (b) is violated if sufficiently light isocurvature degrees of freedom couple to the adiabatic mode during inflation: isocurvature fluctuations ($\delta\sigma$) that do not immediately decay after Hubble crossing would indeed feed super-horizon curvature perturbations \cite{Gordon:2000hv}, $\dot{\zeta}_{k}\sim (\dot{\theta}/H)\delta\sigma_{k}$ for $k\ll aH$, with $ \dot{\theta}/H$ quantifying the magnitude of the curvature-isocurvature coupling. A non-zero coupling corresponds to a curved trajectory in field space: the adiabatic (parallel) and entropic (perpendicular) directions are expected to mix when the trajectory performs a turn. With a straight trajectory, a multi-field model would generally fall in the single-clock cathegory, with indistinguishable predictions from those of SFSR for correlators in the squeezed limit.       \\

 Another instance in which (b) does not apply is when modes cross outside Hubble during a non-attractor fast-roll, often also dubbed as ``ultra-slow'' roll, phase (see e.g. \cite{fastroll}). The simplest example is that of an inflaton rolling on a potential that is transiently flat. In this case, one has $\ddot{\phi}+3H\dot{\phi}=0$, which implies $\epsilon\simeq\dot{\phi}^{2}\propto a^{-6}$ and $\eta\equiv\dot{\epsilon}/\epsilon H\simeq  -6$. As a result, the curvature fluctuation on super-horizon scales undergoes a rapid growth during this stage, $\zeta\sim \epsilon^{-1/2}\propto a^{3}$ \cite{Namjoo:2012aa}. Another simple case is that of an inflaton rolling upward for a few e-folds along its potential (non-attractor stage, with $\epsilon$ rapidly decreasing over time and $\eta\simeq \mathcal{O}(1)$) and then rolling back downward (attractor stage) \cite{Chen:2013aj}. As a result, for modes that exit during the non-attractor stage, a physical correlation may be generated among long and short wavelengths. \\

 Condition (a) is violated in a number of set-ups. A very important case is that of a modified initial state for primordial perturbations. The standard choice is given by Bunch-Davies initial conditions: deep inside the horizon the space-time curvature is negligible in comparison with the scale defined by the wavenumber ($k/a\gg H$), the vacuum state should therefore be equivalent to the Minkowski vacuum. A non-Bunch Davies initial state is not an unreasonable expectation for inflation. First off, inflation is normally formulated in terms of an effective low-energy description of the early Universe, only valid up to a cut-off energy scale. A given pre-inflationary dynamics can leave its imprint in the form of an excited initial state for inflation. Some examples include a non-attractor \cite{Lello:2013mfa} or anisotropic pre-inflationary phase \cite{Dey:2011mj} or loop-quantum cosmology scenarios \cite{Agullo:2013ai}. If the initial state is excited, a physical correlation can arise among hard and soft modes before the latter crosse outside Hubble.\\
 
 Another mechanism that violates (a) by triggering the excitation of sub-horizon fluctuations involves the generation of resonances between perturbations and oscillations in the background \cite{Chen:2008wn}. An oscillating background introduces a new scale represented by its frequency, $\omega_{\text{bck}}$. A distinct phenomenology arises when the latter is larger than Hubble, $\omega_{\text{bck}} \gg H$. The frequency of fluctuations redshifts with time, $\omega_{\text{pert}}(t)=k/a(t) $, therefore modes starting off with a frequency smaller than the background frequency $\omega_{\text{bck}}$ eventually reach $\omega_{\text{pert}}(t_{*})\simeq \omega_{\text{bck}}$. At time $t_{*}$ the system is then driven to oscillate with a much greater amplitude. \\
 \indent Several models have been proposed in this direction. In the context of axion inflation\footnote{Inflationary models endowed with axions have been introduced with the purpose of ensuring that the flatness of the inflationary potential is protected against large renormalization thanks to a nearly exact axionic shift symmetry. This is especially important for large field inflation, where it is more challenging to protect the smallness of the slow-roll parameters over a large enough range in field space.}, for instance, a sinusoidal contribution to the potential is expected from instanton effects (see e.g. \cite{Flauger:2010ja}). Interestingly, in \cite{Flauger:2013hra} it is shown how, in the presence of a potential with a sinusoidal correction, a sub-horizon curvature fluctuation initially in a Bunch-Davies state transitions to an excited state as soon as its frequency matches that of the potential. Many other examples of the sort have been proposed in string theory contexts (see e.g. \cite{Flauger:2009ab}, and \cite{Battefeld:2013xka} for a string-theory inspired multi-field inflation) and in brane inflation (as e.g. in \cite{Bean:2008na}). \\

%Another example of a smooth potential with superimposed periodic modulations was discussed in \cite{1505.01834 pag 39} and in \cite{204} for resonances in a multi-field scenario.\\

 It is also important to stress that single-clock ccs are formally derived from invariance of the Lagrangian under space-diffeomorphisms \cite{maldacena}. When the latter are broken, one expects a multi-clock dynamics. A representative case is solid inflation \cite{Endlich:2012pz}, which is driven by an expanding solid. In contradistinction with most inflationary scenarios (certainly all those captured by the \textsl{effective theory of inflation} \cite{EFT}), in solid inflation time-diffeomorphism invariance is preserved whereas space-diffs are broken. Homogeneity and isotropy are restored by the introduction of internal symmetries for the solid, but the phenomenology of the model is very distinct and can lead, among other signatures, to observable correlations in the squeezed limit (see also \cite{Bartolo:2015qvr} for an effective theory set-up where both time and space diffs are broken).

\subsection{Realizing a strong scale-dependence for the bispectrum}

Scale-dependence of cosmological fluctuations is an important probe of the inflationary dynamics. In standard SFSR inflation, curvature fluctuations are entirely determined by the evolution of the inflaton background. Modes freeze out on super-horizon scales, as a result their amplitude at late times is well-approximated by their amplitude at horizon crossing-time. Different $k$-modes cross outside the Hubble radius at different times, however their amplitude is predicted to remain nearly constant across the inflationary scales because the background is in a slow-roll evolving regime.  \\

 In the presence of a change in the dynamical evolution during inflation occurring at given scales or within a range of scales, the scale-dependence of cosmological fluctuations may be affected at the linear or also at the non-linear levels. This can happen in a variety of scenarios. \\
\indent One possibility is that the slow-roll evolution is disrupted by a step in the potential \cite{step}. Disruption (or interruption) of slow-roll may also occur during phase transitions, as in a number of grand unified theories and in string theory motivated inflationary models \cite{transitions}. \\
\indent  Features and bumps can appear in correlation functions at characteristic scales if massive ($m_{\text{iso}}\simeq H$ or larger) fields were active during inflation at the background level or if they also contributed to cosmological fluctuations. In the case of a turning trajectory in field space, massive isocurvature modes couple to the adiabatic mode and they affect curvature fluctuations in ways that depend on the mass of the heavy fields and on the characteristics of the turn. \\
\indent If the trajectory has a sudden turn, for example, features and bumps in two and higher-order correlation functions are predicted to appear around a characteristic scale correlated with the time at which the turn happens \cite{suddenturn}. On the other hand, a constant turn causes no scale-dependence in the power spectrum, although it can potentially affect the one of higher order correlation functions, depending on the specific set-up. Under certain adiabaticity assumptions, and for very heavy ($m_{\text{iso}}\gg H$) fields, the system is effectively equivalent to single field model where the inflaton has a speed of sound undergoing a transient reduction \cite{transientreduction}. If the isocurvature field mass is instead in the intermediate range ($m_{\text{iso}}\simeq H$), one falls into the class of quasi-single-field inflation, interpolating between the single and the multi-field regimes. In this case an effective single-field theory description is no longer accurate, the predictivity of single-field models is nevertheless preserved since the isocurvature degrees of freedom are not very long lived on super horizon scales. In the case of a constant-turn trajectory, a standard scale-dependence is predicted for the power spectrum: the effect of the heavy fields in this case is that of a mere rescaling of the power spectrum amplitude. A non-standard scaling with momenta in the squeezed-limit is instead predicted in this case for the bispectrum if the masses of the isocurvature modes are not much heavier than Hubble \cite{qsforiginal}. \\
\indent Another interesting inflationary set-up, often motivated by string theory, is that of an inflaton coupled to fields with time-varying masses that become very light at some point during the inflationary evolution, as in \cite{trappedlike}. As a result, bursts of particles arise that can affect the background evolution. Particle production in these models indeed normally occurs at the expense of the inflaton kinetic energy, thus ``slowing down'' the rolling along the inflationary potential. The produced particle can not only back-react on the background evolution itself, but also affect the fluctuations with the appearance of bumpy features in correlation functions. \\  

 Other general set-ups leading to the appearance of features at characteristic length scales in the correlation functions include inflation with modified initial states \cite{nonbd} and models where resonances occur between quantum fluctuations and oscillations in the background, the general mechanisms and motivations for which were briefly reviewed in Sec.~\ref{sec31}.

\subsection{Some representative cases combining all the desirable features}

Models predicting a signal that may be observable in the near future through distortion-temperature anisotropies correlations should predict a physical coupling among hard and soft modes, in addition to a non-Gaussianity amplitude $f_{\text{nl}}$ that grows on small scales. The growth may be monotonic (in this case the signal would be enhanced on configurations \textsl{b} and, even more, \textsl{c} of Fig.~\ref{FIG1}, in comparison with \textsl{a}) or may receive most of its contribution from intermediate scales (configuration \textsl{b}). In the latter case probing the distortion-temperature correlation on $y$-scale would be even more important as those may be the scales where the signal is more strongly enhanced. In the following, we discuss some classes of models equipped with these desirable features.  \\

\noindent \textsl{Curvaton-type models}. \\
This is a class of inflationary models characterized by the presence of one or more ``curvatons'', light ($m_{\text{curv}}\ll H$) weakly-coupled spectator fields. After the decay of the inflaton, the Hubble rate quickly drops over time until it falls below the value of $m_{\text{curv}}$. At that point the curvaton behaves like non-relativistic matter, oscillating and eventually decaying into radiation. During the oscillatory phase, the total energy density contributed by the curvaton background decreases over time more slowly than radiation; as a result the curvaton contribution to the total energy density of the Universe may become sizable by the time it decays. Upon decay, its isocurvature fluctuations ($\delta\sigma$) are converted into curvature fluctuations ($\zeta$), contributing by an amount proportional to the fraction $r\approx [\rho_{\text{curv}}/\rho_{\text{total}}]_{\text{decay}}$ of the curvaton energy density at the time of decay. \\
\indent Depending on the model, the contribution from inflaton field fluctuations may be subdominant (``pure curvaton''), e.g. if the energy scale of inflation is sufficiently low, or non-negligible (``mixed curvaton''). Local-type non-Gaussianity arises from the non-linear evolution of the curvaton fluctuations. For the pure curvaton case one finds \cite{curvaton}
\begin{equation}
\zeta_{\vec{k}} = \frac{2}{3}r\frac{\sigma^{'}_{\text{osc}}}{\sigma^{}_{\text{osc}}}\delta\sigma_{\vec{k}}(t_{k})+\frac{1}{9}\left[3 r\left(1+\frac{\sigma^{}_{\text{osc}}\sigma^{''}_{\text{osc}}}{\sigma^{'2}_{\text{osc}}}\right)-4 r^2-2 r^3\right]\left(\frac{\sigma^{'}_{\text{osc}}}{\sigma^{}_{\text{osc}}}\right)^{2}(\delta\sigma\star\delta\sigma)_{\vec{k}}(t_{k})+\mathcal{O}(\delta\sigma^3)\,,
\end{equation}
where $\sigma_{\text{osc}}$ is the background value of the curvaton field at the beginning of the oscillation and $t_{k}$ is the horizon crossing time during inflation. Notice that the non-Gaussian part depends more strongly than the Gaussian part on the dynamics of the background field. From here, one finds
\begin{equation}\label{sdngcurvaton}
f_{\text{nl}} =\frac{5}{4r}\left(1+\frac{\sigma_{\text{osc}}\sigma^{''}_{\text{osc}}}{\sigma^{'2}_{\text{osc}}}\right) -\frac{5}{3}-\frac{5r}{6}\,.
\end{equation}
The scale dependence of $f_{\text{nl}}$ is due to the dependence of $\sigma_{\text{osc}}$ from the background value of the curvaton at $t_{k}$. The potential of the curvaton can be generically written as a quadratic plus a self-interaction contributions, $V(\sigma)=(1/2)m_{\sigma}^{2}\sigma^2+\lambda m_{\sigma}^{4}(\sigma/m_{\sigma})^{n}$. The closer the curvaton is to the time of decay, the smaller its field values and the safer it is to approximate the potential with its quadratic contribution. At earlier times though, as one can see from Eq.~(\ref{sdngcurvaton}), self-interactions have important phenomenological implications\footnote{Self-interactions can often arise from curvaton models embedded in MSSM set-ups \cite{MSSM}. They are also expected to arise at loop level even when absent at tree-level: even though weakly coupled, the curvaton will eventually decay and therefore it is expected to be coupled to other fields \cite{Byrnes:2010xd}.} (in the quadratic case one finds $\sigma_{\text{osc}}\sigma^{''}_{\text{osc}}/\sigma^{'2}_{\text{osc}}=0$). The effect due to self-interactions is then more important in the $r\ll 1$ limit. \\
\indent The specific scale-dependence of the non-Gaussianity is very much dependent on the given realization and model parameters, including the form of the self-interacting potential (for $n\geq 6$, for instance, $f_{\text{nl}}$ undergoes oscillations as a function of $\sigma(t_{k})$), leading to an amplitude $|f_{\text{nl}}|$ that can vary strongly with scales (see also e.g. \cite{Byrnes:2015asa}).\\

\noindent \textsl{More models with fields endowed with time-varying masses}. \\
In \cite{Riotto:2010nh} it was shown that in the presence of a spectator field that is heavy ($m_{\sigma}\gg H$) up to a given scale $k_{*}$ and lighter afterwords, one finds a sizable local non-Gaussianity for modes crossing outside Hubble during the second phase and a negligible one during the first phase. A local non-Gaussianity in this set-up is sourced by the non-linearities in the spectator field fluctuations. For modes exiting during the first phase, ($t\leq t_{*}$), the spectator field fluctuations are suppressed by an $H/m_{\sigma}$ factor; in additions to this, they redshift like matter ($\delta\sigma\sim a^{-3}$). When the mass becomes small ($t\gg t_{*}$), these modes nearly freeze to a value 
\begin{equation}\label{onet}
| \delta\sigma_{k}(t_{\text{end}}) |^2\sim \frac{H^2}{k^3} \frac{H}{M} \left(\frac{k}{a_{*}H}\right)^3 \left(\frac{a_{*}H}{a_{\text{end}}H}\right)^{3-2\nu_{m}}\,, 
\end{equation}
where $\nu_{m}^{2}\simeq 3/2-m^2/(3H^2)$ and $t_{\text{end}}$ is the end of inflation. Modes exiting during the second phase, between $t_{*}$ and the end of inflation, freeze on superhorizon scales immediately and their super-horizon fluctuations at the end of inflation reads 
\begin{equation}\label{twot}
|\delta\sigma_{k}(t_{\text{end}})|^2\sim  \frac{H^2}{k^3}\left(\frac{k}{a_{\text{end}}H}\right)^{3-2\nu_{m}}\,.
\end{equation}
In the following phase (phase three, $t_{\text{end}}\leq t \leq t_{\text{dec}}$), the Hubble radius (initially larger than $m_{\sigma}$) decreases quickly over time up to the point where it becomes smaller than the spectator field mass again, causing it to behave as non-relativistic matter and eventually decay (at $t_{\text{dec}}$). The power spectrum of the spectator field for $t_{\text{end}}\leq t \leq t_{\text{dec}}$ is then given by (\ref{onet}) or (\ref{twot}) (depending on when the specific mode exited during inflation), multiplied by an extra $\sim (a/a_{\text{end}})^3$ factor to account for the damping during oscillations. The total curvature fluctuation after the inflaton has decayed into radiation is given by $\zeta\approx \zeta_{\text{rad}}+\delta\rho_{\sigma}/4\rho_{\text{tot}}$ where $\delta\rho_{\sigma}/4\rho_{\text{tot}}=(m^2/2)|\delta\sigma_{\vec{k}}|^2/\rho_{\text{tot}}$. \\
\indent If the non-Gaussianity in the squeezed limit from the inflaton field is negligible (this is certainly the case if, for instance, the couplings between inflaton and spectator field during inflation are very weak), the main contribution to the squeezed bispectrum comes from the spectator field modes exiting during the second phase, amounting to an amplitude \cite{Riotto:2010nh}
\begin{equation}
f_{\text{nl}}\approx 2\times  10^{-3}\eta_{\sigma}^{1/2}\mathcal{P}_{\zeta}^{-1/2}\,,
\end{equation}
with $\eta_{\sigma}\equiv 2m^2/3H^2$. \\
\indent A crucial difference between this and the curvaton model is that the spectator field transitions from a heavy to a light regime during inflation, whereas in the curvaton case the field is light throughout inflation. A consequence of this choice is also that $\delta\rho_{\sigma}\sim \delta\sigma$ for the curvaton, whereas $\delta\rho_{\sigma}\sim \delta\sigma^2$ in \cite{Riotto:2010nh} (the background value of the curvaton quickly decreases in time during the first, massive, phase). An example for a concrete realization of this set-up includes an inflaton field ($\phi$), a scalar spectator $\sigma$ and an additional scalar field $\chi$, with a potential \cite{Riotto:2010nh} 
\begin{equation}
V=V(\phi)+\frac{1}{2}\left(g^2 \chi^2+m^2\right)\sigma^2+\frac{1}{2}\left(-m_{\chi}^{2}+h^2\phi^2\right)\chi^2+\frac{1}{4}\lambda\chi^4\,.
\end{equation}
With this potential, one can have a two-stage inflation where $\chi$ is initially frozen to $\langle \chi\rangle^2=m_{\chi}^2/\lambda$, which implies an effective mass $M^2\sim(g^2m_{\chi}^{2}/\lambda)$ for $\sigma$ (this can be chosen to be larger than $H^2$). When $\langle \phi\rangle^2$ becomes larger than $m_{\chi}^{2}/h^2$, $\langle\chi\rangle\simeq 0$ and the mass of the spectator field becomes equal to $m^2$, which is assumed to be small during inflation. \\

\noindent \textsl{Sdng from isocurvature-inflaton interactions}\\
In the two classes of models just described, local non-Gaussianity is only generated from the isocurvature field after inflation, the couplings between curvaton and inflaton being small during inflation; another possibility is that those couplings are instead important. \\
\indent One could for instance think about a scenario where a QsF-type (in the simplest case this includes an inflaton and an isocurvature degree of freedom with a mass $m_{\text{iso}}\approx H$) phase is preceded by an effectively single-field phase. This would be the case if the trajectory in field space is straight at first, and performs a turn later on. For modes that cross outside the Hubble radius during the phase characterized by a curved trajectory one expects a non-zero correlation in the squeezed limit. For relatively small values of the isocurvaton mass, $m_{\text{iso}}\lesssim 3H/2$, and for a soft turn, the predictions for the squeezed bispectrum in the minimal QsF model \cite{qsforiginal} would likely be a good approximation at those scales that cross outside Hubble during the turning phase
\begin{equation}\label{yy}
f_{\text{nl}}\simeq \alpha(\nu)\mathcal{P}^{-1/2}_{\zeta}\left(-\frac{V^{'''}}{H}\right)\left(\frac{\dot{\theta}}{H}\right)^{3}\,,
\end{equation}
where $\alpha$ is a function of the isocurvaton mass ($\nu\equiv\sqrt{9/4-m^2/H^2}$), $\dot{\theta}$ is the angular velocity along the trajectory in field space (here approximated as a constant for the whole duration of the turn) and $V$ is the potential of the isocurvature field. Remarkably, the amplitude in (\ref{yy}) can be sizable since the isocurvature field, unlike the inflaton, needs not be in slow-roll (all that is required is that $(V^{'''}/H)\lesssim 1$ for the perturbative expansion to hold). \\
\indent An exact treatment would also need to account for the behavior of the fluctuations at the point where the turn begins: here one expects some superimposed features in the correlation functions (these would be an additional signature one would be searching for in the data). In the case of a slow turn, the features would only appear on a very limited range of scales near the turning point; as a result, if the turn last for a sufficient number of e-folds one may be able to safely neglect the effects of the initial bending of the trajectory. One could envisage the case where the trajectory becomes curved on scales of $0.1 \text{Mpc}^{-1}$ or smaller, thus incorporating the smallest scales we can detect from CMB anisotropies along with the $y$ and $\mu$-distortion scales \footnote{Notice that for a physical correlation to arise in the squeezed limit from isocurvature modes as in QsF inflation, the long-wavelength mode needs to be sourced by the isocurvature mode, as a result we only expect a squeezed-limit signal if the long-wavelength mode crosses outside the horizon when the trajectory begins to turn.}.\\

\indent We have so far discussed the possibility of generating sdNG in the squeezed limit with an enhancement on smaller scales in a variety of multi-field classes. If the additional fields are heavier than Hubble and behave as an oscillating background during inflation, a sdnG can arise from the resonance between the oscillating background and the inflaton perturbations on sub-horizon scales (a phenomenon of limited duration in time). We will consider this possibility in Sec.~\ref{resonance}, where we show how in this case the squeezed signal from distortion-temperature correlations can be large on $y$ scales (as opposed to $\mu$ scales). In Sec.~\ref{g-hybrid} we present a model where $f_{\text{nl}}$ grows monotonically towards smaller scales and the strongest enhancement is therefore expected on $\mu$ scales.

\section{Resonant features from heavy fields}
\label{resonance}

As discussed in Sec.~\ref{general2}, in the presence of an oscillatory background with a frequency $\omega_{\text{bck}}$, fluctuations with frequency $\omega_{\text{pert}}\sim k/a$, larger than $\omega_{\text{bck}}$ at the onset of the oscillations, eventually reach a point, during inflation, where their frequency resonates with the external one, causing the amplitude of fluctuations to undergo a transient amplification. For this effect to be appreciable, the frequency of the background oscillations should be much larger than Hubble. This implies that the mode numbers of fluctuations that are affected by the resonance are sub-Hubble as well. This resonance effect can be interpreted as an excitation of sub-horizon fluctuations, which leads to the expectation that correlations are produced in these models even before modes reach Hubble size, thus also allowing for a correlation among modes of different wavelengths to arise before the long-wavelength mode becomes super-horizon and freezes. The resonance is generally a transient effect and a non-trivial correlation among different wavelengths can only arise in finitely squeezed configurations, limited by the time-scale of the oscillation.\\

 In \cite{Saito} resonances are signatures of heavy fields ($m\gg H$). The fields oscillate with a frequency proportional to their mass and eventually decay, producing sharp peaks in the correlation functions at some given scales. \\
\indent In the simplest case of a bi-dimensional field space, with an inflaton ($\phi$) and a heavy field ($\chi$), and endowing the inflationary Lagrangian with an approximate shift symmetry in order to protect the flatness of the potential from large quantum corrections, it is natural to include derivative couplings of the inflaton \cite{Saito} and the action is given by
\begin{equation} \label{euuu}
S = - \int d^4x \sqrt{-g} \bigg{[} \frac{1}{2}(\partial \phi)^2 + \frac{1}{2}(\partial \chi)^2 + V(\phi) + \frac{m_{\chi}^2}{2}\chi^2 + K_n + K_d + K_p + O(X \phi^2/\Lambda_n^2, X^3/ \Lambda^8_d )
\bigg{]}\,,
\end{equation} 
where $X\equiv-(\partial\phi)^2$ and 
\begin{eqnarray}
K_n &=& \frac{\lambda_n}{2 \Lambda_n} \chi \partial \phi^2\,, \\
K_d &=& \frac{\lambda_{d1}}{4 \Lambda_d^4}(\partial \chi)^2 (\partial \phi)^2 + \frac{\lambda_{d2}}{4 \Lambda_d^4}(\partial \chi\cdot\partial \phi)^2 \,,\\
K_p &=& \frac{\lambda_{p}}{4 \Lambda_d^4}(\partial \chi\cdot\partial \phi)(\partial \phi)^2\,.
\end{eqnarray} 
$K_{n}$ and $K_{d}$ arise at leading order in $\Lambda_{n}$ and $\Lambda_{d}$ if one assumes both a parity, $\phi\rightarrow -\phi$, and a shift symmetry, $\phi\rightarrow \phi+c$, and enforces the conditions $\chi\ll\Lambda_{n}$ and $\dot{\phi},\,\dot{\chi}\ll\Lambda_{d}^{2}$ at the background level. If the Lagrangian is not invariant under parity, interactions as $K_{p}$ are also permitted.  \\ 
\indent Derivative couplings have been shown \cite{Saito} to strongly affect the evolution of the adiabatic fluctuations through the resonance, while at the same time producing a negligible effect on the inflationary background because of the slow-roll evolution of the inflaton field. The caveat behind this finding is that the contribution of the heavy field to the total energy density is negligible w.r.t. the inflaton potential. The latter condition translates into a condition $\dot{\chi}^2+m^2\chi^2\ll M_{P}^{2}H^2$, where $M_{P}=2.4 \times 10^{18}\,\text{GeV}$ is the reduced Planck mass. \\
\indent The effects of $K_n$ and $K_d$ on the scalar bispectrum have been accounted for in details in \cite{Saito}, although not in the context of spectral distortion-temperature correlations. In what follows, we will focus primarily on $K_p$. The enhancement coming from the latter is indeed even more interesting than $K_{d,n}$ because it is larger: for both $K_n$ and $K_d$ cubic self-interactions for the inflaton field are mediated by gravity and therefore expected to be subleading w.r.t. the parity violating term, for which there is a direct cubic interaction (see Appendix~A for a comparison among the various contributions).

\subsection{Background evolution and power spectrum}

The background and the power spectrum analysis were performed in \cite{Saito}. With the conditions previously defined for the background fields, $\chi\ll\Lambda_{n}$ and $\dot{\phi},\,\dot{\chi}\ll\Lambda_{d}^{2}$, and introducing the slow-roll parameters $\epsilon_{V}\equiv (M_{P}^{2}/2)(V^{'}/V)^{2}\ll 1$ and $|\eta_{V}|\equiv M_{P}^{2} V^{''}/V\ll 1$, the background equation of motion for the inflaton is well-approximated by the standard $3 H \dot{\phi}\simeq -V^{'}$. \\
\indent The heavy field background evolves in time as $\chi(t)\simeq \chi_{0} e^{-\Gamma t}\cos(mt)\theta(t-t_{0})$, where the decay constant is assumed to be $H\ll \Gamma\ll m$, and  $\theta(t-t_{0})$ is the Heaviside function (having defined $t=t_{0}$ as the onset of the oscillation). For any given (physical) momentum $k/a(t)$ larger than the external frequency $m$ at the onset of the oscillation, $k\gg m$, there comes a time $t_{*k}$ such that $k/a(t_{*k})\approx m$. The horizon crossing time for the same mode is defined by the condition $k/a(t_{k})= H$. Since by definition $m\gg H$, we have $a(t_{*k})\ll a(t_{k})$, i.e. the resonance can only affect subhorizon modes. \\

\begin{figure}
	\begin{center}
	\includegraphics[width=0.6\textwidth]{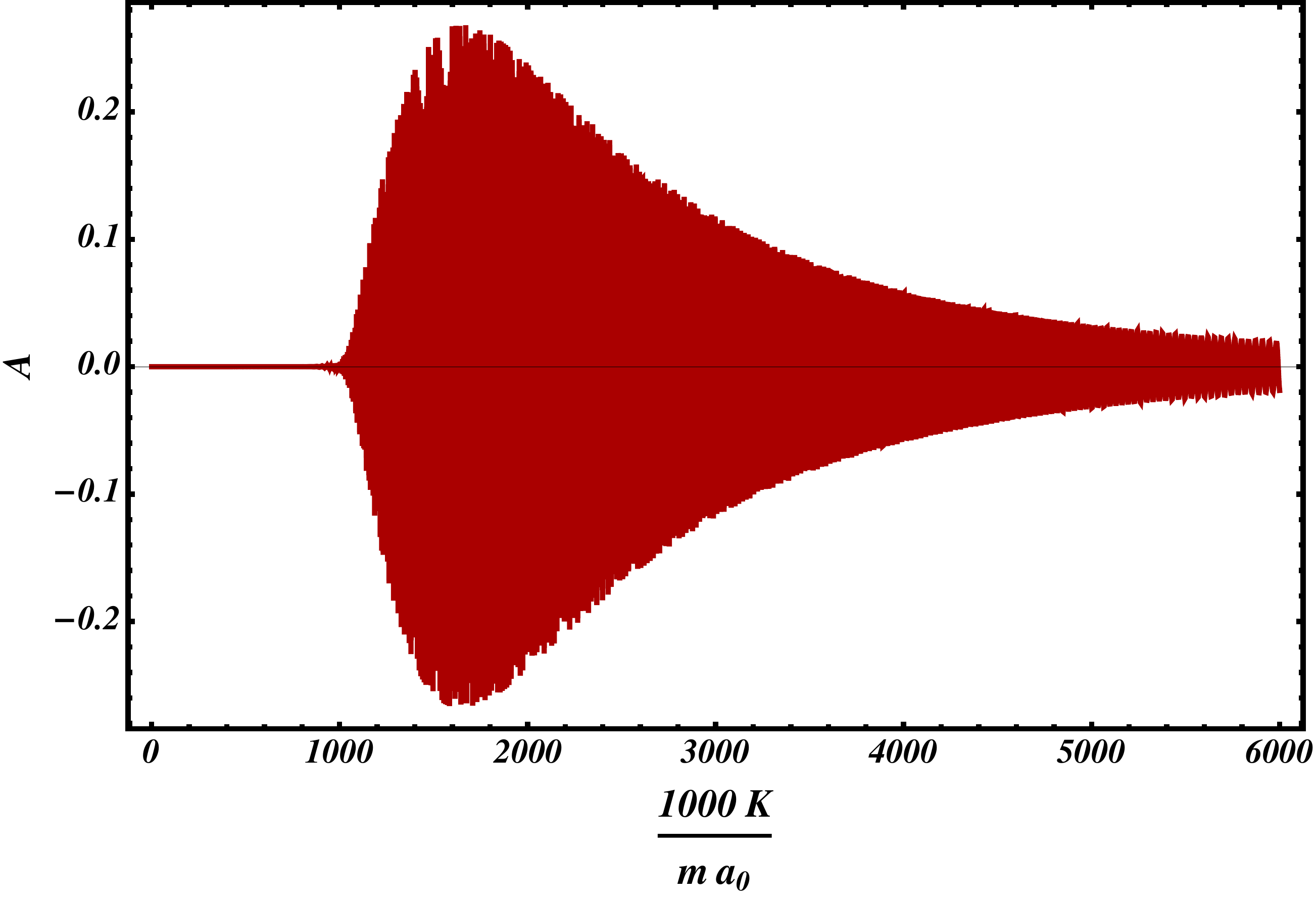}
	\caption{Representation of the integral in Eq.~(\ref{ttt}) for the bispectrum, with a peak appearing around $K\simeq m a_{0}$, with $K=k_1+k_2+k_3$.}
		\label{FIG2}
	\end{center}
	\end{figure}

\indent The correction to the amplitude of the scalar power spectrum from vacuum fluctuations due to the interactions in the Lagrangian has the following form
\begin{equation}\label{cccc}
\Delta\langle\delta\phi_{\vec{k}_{1}}(t)\delta\phi_{\vec{k}_{2}}(t)  \rangle\simeq i\int_{0}^{t} dt'\langle\left[H_{I}^{(2)}(t'),\delta\phi_{\vec{k}_{1}}(t)\delta\phi_{\vec{k}_{2}}(t)\right] \rangle\,,
\end{equation}
where the second order Hamiltonian $H_{I}^{(2)}$ is obtained from the expansion of the $K_{i}$ interaction terms ($i=n,d,p$) in (\ref{euuu}) \cite{Saito}. The derivative couplings can be safely treated as a correction to the free Lagrangian if the couplings $q_{n}\equiv\lambda_{n}\chi_{0}/\Lambda_{n}$, $q_{dj}\equiv\lambda_{dj}m^2\chi_{0}^{2}/(2\Lambda_{d}^{4})$ ($j=1,2$) and $q_{p}\equiv\lambda_{p}m\chi\dot{\phi}/(2\Lambda_{d}^{4})$ are assumed to be small. Contributions as (\ref{cccc}) can therefore be appreciably large only in the vicinity of those wave-numbers for which (\ref{cccc}) has a peak due to resonances. It is straightforward to show (see also Appendix A for details) that, as a result of the resonance between the heavy field background and the fluctuations in the inflation field, the amplitude of the power spectrum exhibits a peak for wave-numbers close to $k_{p}\simeq m a_{0} e^{\sqrt{H/m}}/2$. In the limit where the decay rate is small, $\Gamma\ll\sqrt{Hm}$, the overall correction to the curvature power spectrum (using $\zeta\sim(H/\dot{\phi})\delta\phi$) takes the simple form
\begin{equation}
\frac{\Delta\mathcal{P}_{\phi}^{(i)}}{\mathcal{P}_{\phi}}\simeq q_{i}\sqrt{\frac{m}{H}}\,,
\end{equation}
where $i=n,d,p$ and $\mathcal{P}_{\phi}\equiv P_{\phi}(k^{3}/2\pi^2)$. As stressed in \cite{Saito}, a few conditions have to be in place to ensure that the perturbative method is valid. These include setting $q_{i}\sqrt{M/H}<1$ and $m<\Lambda_{i}$, the latter stemming from the requirement that loop corrections originating from the derivative interactions do not exceed the tree-level nor the resonance contributions from the same interactions.

\subsection{Squeezed limit bispectrum}

The resonance of the inflaton fluctuations with the oscillating massive field gives rise to an amplification of the bispectrum amplitude when $K\equiv k_{1}+k_{2}+k_{3}\simeq k_{p}$ \cite{Saito} (see also Fig.~\ref{FIG2} and Appendix~A for more details). The expression for the scalar bispectrum was computed for $K_{n}$ and $K_{q}$ in \cite{Saito}. In the following we compute the contribution from parity violating derivative interactions such as $K_{p}$. Using the in-in formula (see e.g. Eq.~(\ref{ininf})) one finds 
\begin{eqnarray}\label{rrr}
B_{p}(k_{1},k_{2},k_{3})&\simeq & \frac{q_{p}}{32}\left(e^{-2\Gamma/\sqrt{mH}}-1\right)\frac{\cos(\theta_{*})}{\epsilon_{V}^{2}}\left(\frac{m}{\Gamma}\right)\left(\frac{m}{H}\right)^{2}\left(\frac{H}{M_{p}}\right)^{4}\nonumber\\&\times&\frac{1}{k_{p}^3 k_{1}k_{2}k_{3}}\left[3-\frac{k_{1}\left(\vec{k}_{2}\cdot\vec{k}_{3}\right)+k_{2}\left(\vec{k}_{1}\cdot\vec{k}_{3}\right)+k_{3}\left(\vec{k}_{1}\cdot\vec{k}_{2}\right)}{k_{1}k_{2}k_{3}}\right]
\end{eqnarray}
where we defined $\langle\zeta_{\vec{k}_{}}\zeta_{\vec{k}_{2}}\zeta_{\vec{k}_{3}} \rangle = (2\pi)^{3}\delta^{(3)}(\vec{k}_{}+\vec{k}_{2}+\vec{k}_{3})B(k_{1},k_{2},k_{3})$ and $\theta_{*}\simeq (m/H)[1+\ln(2k/m)]$ is the phase function at the resonance. This expression has been derived under the assumption that the three modes are all sub-horizon around $t_{*}$, defined by $K/a(t_{*})\simeq m$.

\begin{figure}
\centering
	\includegraphics[width=0.9\textwidth]{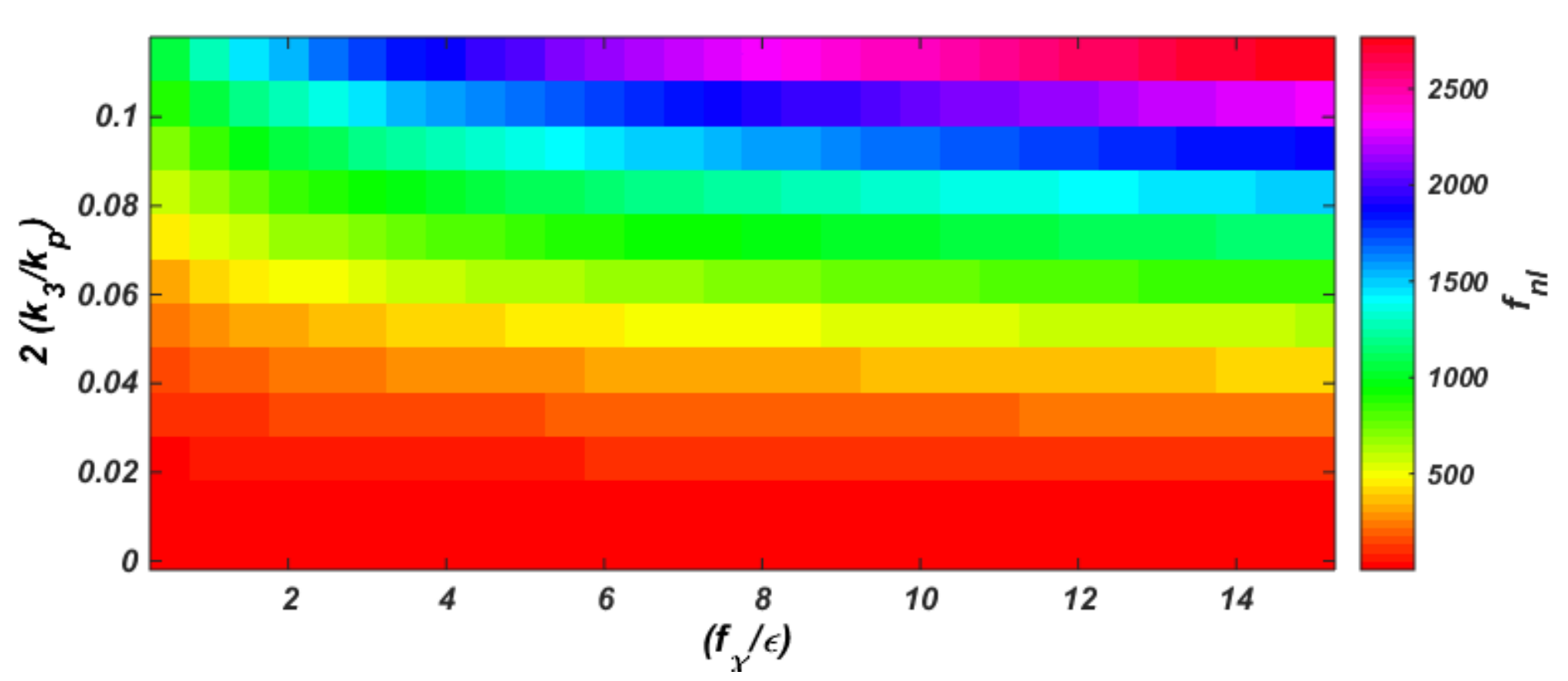}
	\caption{Representation of $f_{\text{nl}}$ values as a function of model parameters, $f_{\chi}/\epsilon_V$, and of the ratio of the CMB ($k_3$) to the distortion scale ($k_{d}\approx k_p/2$). The range of $k_{3}/k_{p}$ has been chosen in such a way as to satisfy the finitely squeezed limit defined in Eq.~(\ref{condf}).}
	\label{FIG3}
\end{figure}

\begin{figure}
	\includegraphics[width=0.45\textwidth]{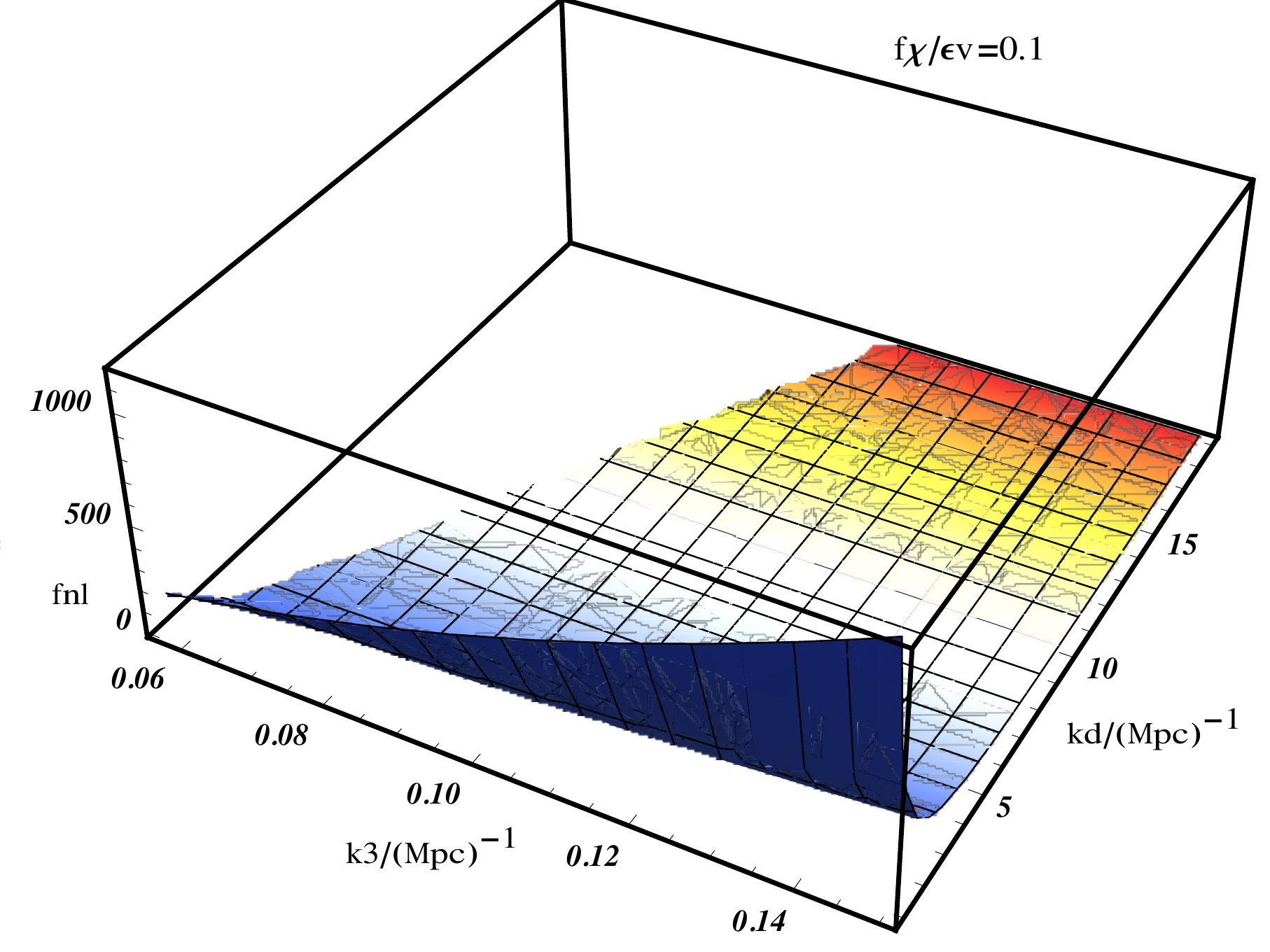}
		\includegraphics[width=0.45\textwidth]{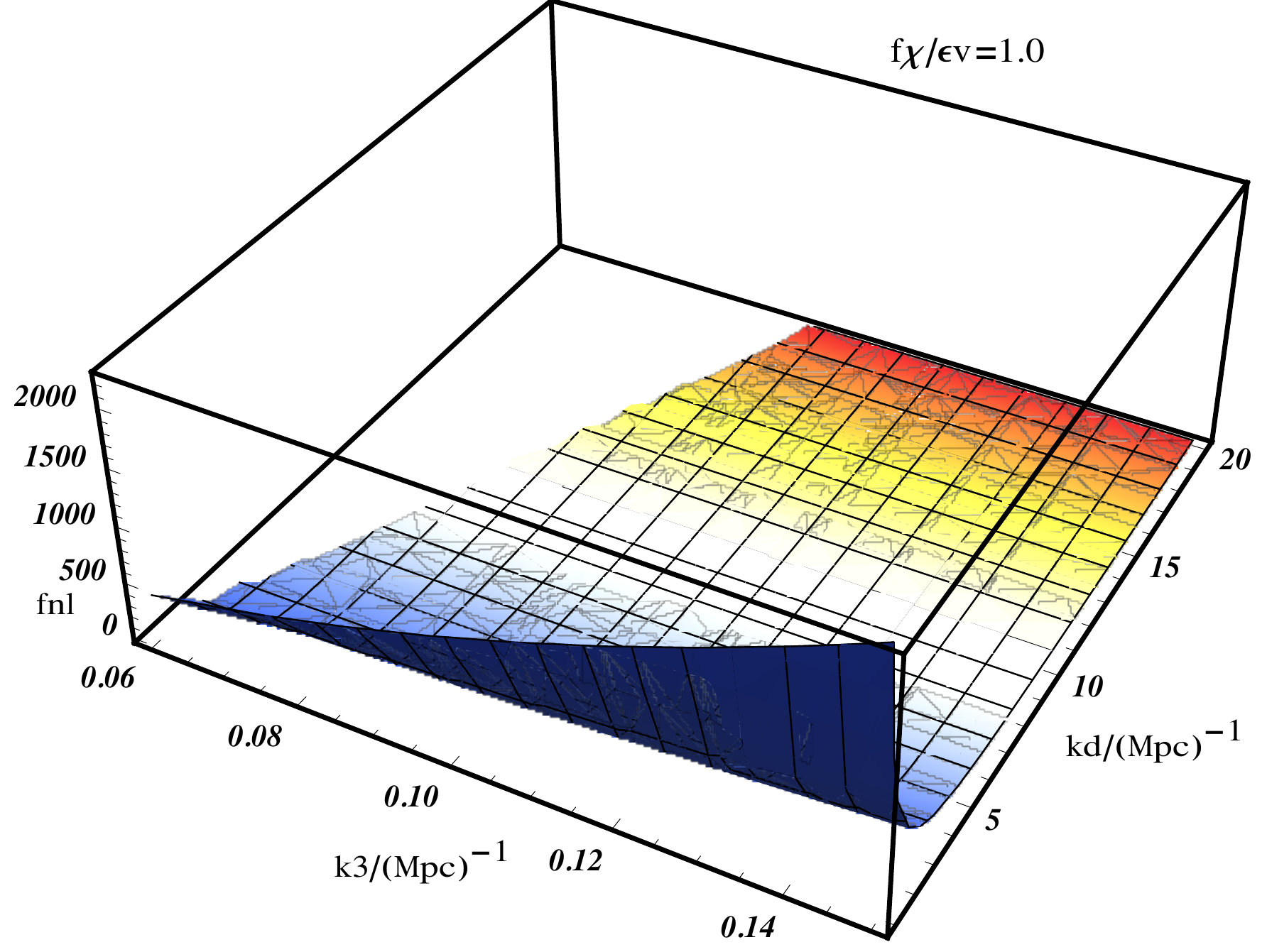}
			\includegraphics[width=0.45\textwidth]{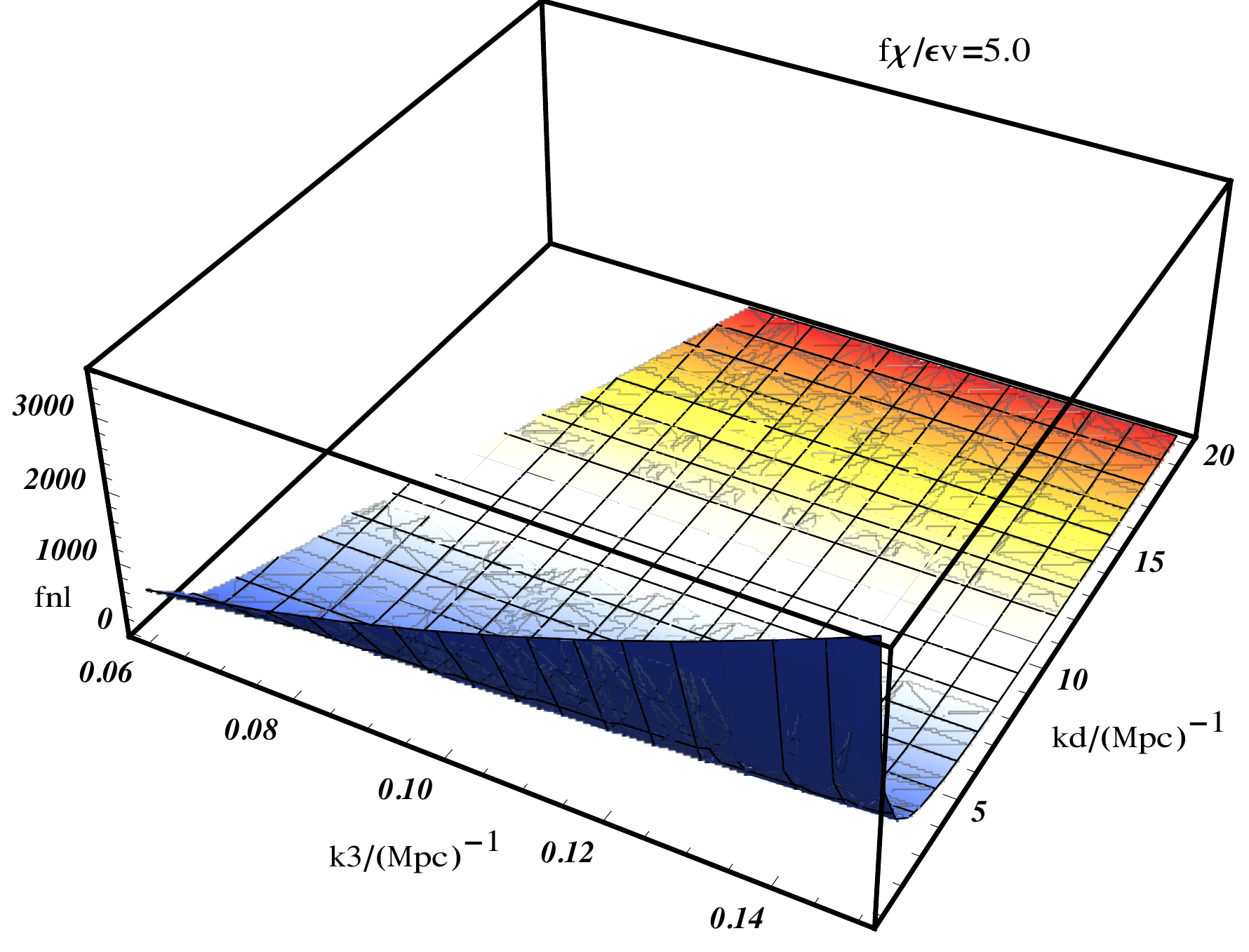}\,\,\,\,\,\,\,\,\,\,\,\,\,\,\,\,\,\,\,\,\,\,\,\,\,\,\,\,\,\,\,
				\includegraphics[width=0.45\textwidth]{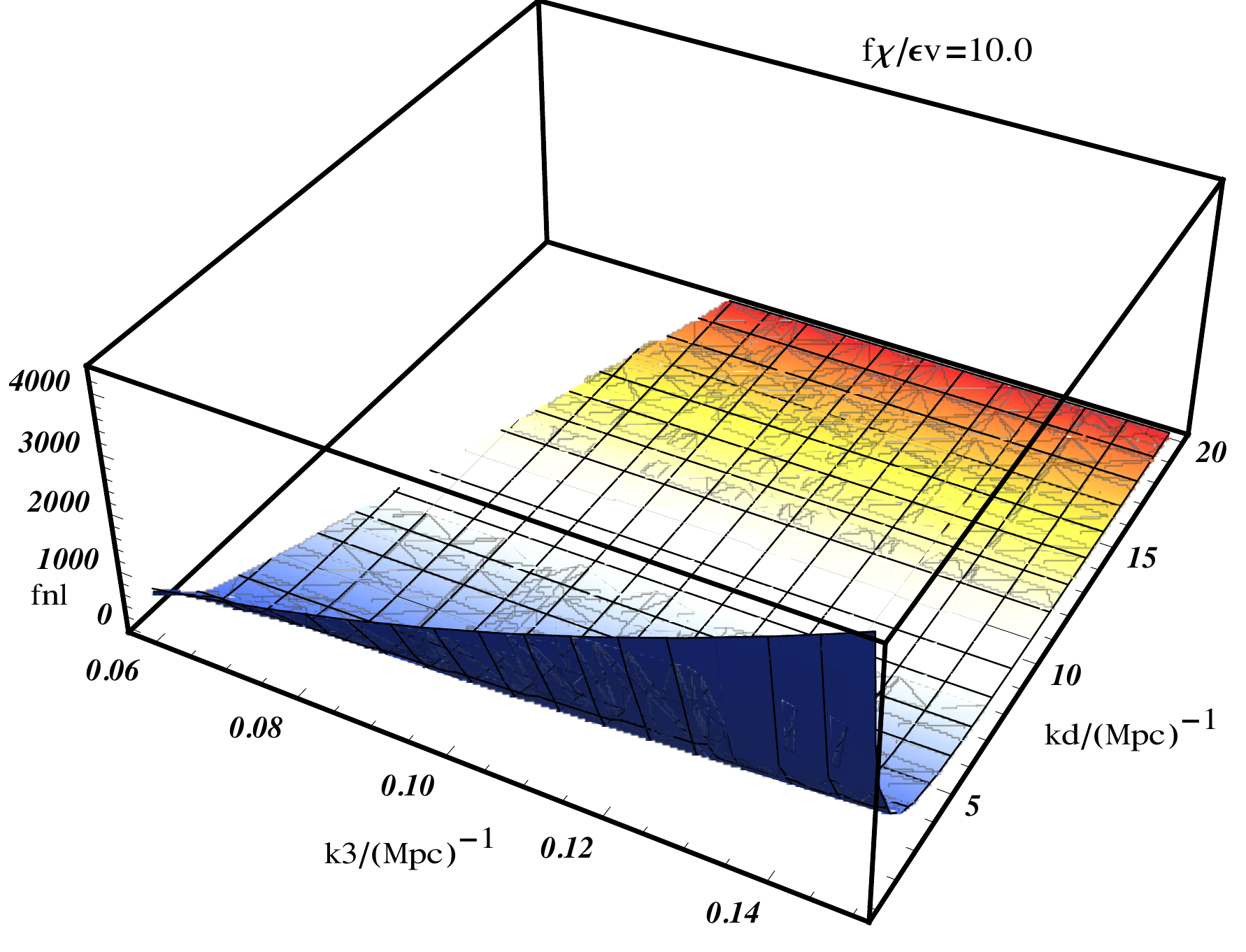}
	\caption{Plot of the squeezed non-Gaussianity amplitude in the $(k_3,k_d)$ plane, for different values of $f_{\chi}/\epsilon_V$.}
	\label{FIG4}
\end{figure}

\noindent The final result for the bispectrum amplitude in the finitely squeezed configuration, with $k_{1}\simeq k_{2}\gg k_{3}>k_{p}(H/m)$ (the lower bound on $k_{3}$ must hold in the range of momenta for which (\ref{rrr}) applies) and $K\simeq k_{p}$, is given by 
\begin{equation}\label{comb3}
f_{\text{nl}}=\left(\frac{5}{12}\right)\left(1-e^{-2\Gamma/\sqrt{mH}}\right)q_{p}\cos\Theta_{*}\frac{m}{\Gamma}\left(\frac{m}{H}\right)^{2}\left(\frac{k_{3}}{k_{p}}\right)^{2}\,.
\end{equation}
For $\Gamma\ll \sqrt{mH}$ one finds
\begin{equation}\label{one}
f_{nl}\simeq \frac{5}{6} q_{p}\left(\frac{m}{H}\right)^{5/2}\left(\frac{k_{3}}{k_{p}}\right)^{2}\,.
\end{equation}
\noindent The power spectrum from the parity violating interaction has the following form
\begin{equation}
\frac{\Delta\mathcal{P}}{\mathcal{P}}\approx  q_{p}\sqrt{\frac{m}{H}}\,.
\end{equation}
Considering the limiting values $\Delta\mathcal{P}/\mathcal{P}\simeq 1$ and $m\simeq 2\pi\Lambda_{d}$, one can show that
\begin{equation}\label{four}
q_{p}\simeq \left(\frac{H}{m}\right)^{4}\sqrt{\frac{f_{\chi}}{\epsilon_{V}}}\left(\frac{(2\pi)^4\sqrt{2}\sqrt{3}}{2\cdot 8\pi^2 \mathcal{P}}\right)\,,
\end{equation}
where $\mathcal{P}\simeq 2.4\times 10^{-9}$. From here we have 
\begin{equation}\label{three}
\frac{m}{H}\simeq \left(\frac{f_{\chi}}{\epsilon_{V}}\right)^{1/7}\left(\frac{(2\pi)^4\sqrt{2}\sqrt{3}}{2\cdot 8\pi^2 \mathcal{P}}\right)^{2/7}
\end{equation}
Notice that this is valid for $k_{3}>(H/m) k_{p}$. The latter condition can be rewritten as
\begin{equation}\label{condf}
\frac{k_{3}}{k_{p}}>\left(\frac{\epsilon_{V}}{f_{\chi}}\right)^{1/7}\left(\frac{(2\pi)^4\sqrt{2}\sqrt{3}}{2\cdot 8\pi^2 \mathcal{P}}\right)^{-2/7}\simeq 10^{-3} \left(\frac{\epsilon_{V}}{f_{\chi}}\right)^{1/7}\,.
\end{equation}
Replacing (\ref{four}) and (\ref{three}) in Eq.~(\ref{one}) one finds
\begin{equation}
f_{\text{nl}}\simeq 4\times 10^5 \left(\frac{f_{\chi}}{\epsilon_{V}}\right)^{2/7}\left(\frac{k_{3}}{k_{p}}\right)^{2}\,.
\end{equation}
Notice that Eq.~(\ref{condf}) places a lower bound of $0.4$ on $f_{\text{nl}}$. Correlations between $\mu$ and $T$ would mostly fall outside the range defined by (\ref{condf}), whereas setting $k_{p}\approx 2k_{y}$, with $k_{y}$ defined on $y$ scales, can lead to values $f_{\text{nl}}\gtrsim \mathcal{O}(10^2)$ or larger. This is shown in Fig.~\ref{FIG3} where we plot the value of $f_{\text{nl}}$ as a function of $f_{\chi}/\epsilon_{V}$ and for different $k_3/k_p$. The results are also shown in $(k_{3},k_{p})$ space, for different values of $f_{\chi}/\epsilon_{V}$ (Fig.~\ref{FIG4}).

\section{Gauged Hybrid Inflation realization of the kinetic coupling}
\label{g-hybrid}

As a second example of models that may predict a strong scale-dependent non-Gaussianity, we consider scenarios with kinetic mixing between the inflaton and another light field. As we will show in the following, a time dependence in the kinetic coupling leads to a scale-dependent $f_{\text{nl}}$. Moreover, this scale dependence is expected to be logarithmic. However, since the running of a logarithmic scale dependence is not that big, we may be unable to create a small enough signal in the CMB temperature anisotropies window, $f_{\text{nl}} \sim O(1)$, while obtaining a big enough signal at small scales (e.g. $f_{\text{nl}} \sim O(100)$ for detection by a PIXIE-like experiment). \\
\indent In order to obtain bispectrum predictions in the desired ranges, one can incorporate such a coupling in a Hybrid realization of inflation. In these models there are two possibilities to generate the curvature perturbation, i.e. during inflation and at the surface of the end of inflation. A cancellation may easily occur between these two contributions at some particular scale, here the CMB scale, while generating a big signal on small scales. \\

 In the following, we consider a multi-field inflationary model where there is a kinetic coupling between two of them and there is also a waterfall field which is coupled to both of the above fields and is responsible to terminate inflation abruptly. As we will see in more detail, the surface of end of inflation in this model is an ellipse in field space and there is a fluctuation associated with the angle of this ellipse. \\
\indent We will also restrict our attention to a vector realization of the additional field. We further assume a complex waterfall field, which is coupled directly to the gauge field.
As we will show, due to the mixture of the gauge kinetic coupling as well as the induced perturbations at the surface of end of inflation, one can obtain an enhancement of $f_{\text{nl}}$ at small scale while being consistent with CMB constraints for the local $f_{\text{nl}}$. Moreover, since we are living in a multi-field set-up, we do expect an enhancement in the squeezed limit non-Gaussianity. \\
\indent  It is important to emphasize that while we chose to focus on a vector realization of the additional light field, one can simply relax this assumption and replace this field with a scalar, thus enlarging this class of models. \\
\indent It is also important to mention that, in order to have a well defined inflationary direction, a nearly scale invariant power in the direction of the light field and a controllable change in the spectral index, the window of allowed functions of the gauged kinetic coupling should be narrowed, with $f(a) \propto a^{-2c}$ where $c$ is very close to $1$: $\left(c-1\right)/c \lesssim 7.5\times 10^{-5} $.

\subsection{Gauged Hybrid Model}
This model is based on the action \cite{Emami:2010rm, Emami:2011yi}
\ba
\label{action3} S=\int d^4 x  \sqrt{-g} \left [ \frac{M_P^2}{2} R - \frac{1}{2} \partial_\mu \phi
\,  ^\mu\phi- \frac{1}{2} D_\mu \psi
\,  D^\mu\bar\psi - \frac{f^{2}(\phi)}{4} F_{\mu \nu} F^{\mu
	\nu}- V(\phi, \psi, \bar \psi) \right]  \, ,
\ea
where $\phi$ denotes the inflaton field and $\psi$ is the complex waterfall field. The covariant derivative is defined by $D_\mu \psi = \partial_\mu  \psi + i \, \mathbf{e} \,  \psi  \, A_\mu $. The gauge field strength is defined  by $F_{\mu \nu} = \nabla_\mu A_\nu
- \nabla_\nu A_\mu  = \partial_\mu A_\nu - \partial_\nu A_\mu $. The gauge kinetic coupling $f(\phi)$ is chosen in such a way as to break the conformal invariance, so that the gauge field survives the expansion at the background and perturbations levels.\\
Since the waterfall field is a complex scalar field, one can write it as
\ba
\psi(x) = \chi(x) \,  e^{i \theta(x)}\,,
\ea
where $\chi(x)$ denotes the radial part and $\theta(x)$ is the angular part. \\
It is convenient to consider the configurations in which the potential has an axial symmetry and $V(\psi, \bar \psi , \phi)= V(\chi, \phi)$. In this way, the potential is just a function of the radial part of the waterfall field and it has the standard hybrid inflation form with the waterfall term being replaced with the radial part of $\psi(x)$ \cite{Linde:1993cn}
\ba
\label{pot3} V(\phi, \chi)=\frac{\lambda}{4 }  \left(  \chi^2 - \frac{M^2}{\lambda} \right)^2 + \frac{g^2}{2} \phi^2 \chi^2 + \frac{m^2}{2} \phi^2 \, .
\ea
Here $\lambda$ and $g$ are dimensionless couplings while $m$ and $M$ has the dimension of mass.\\
\indent By choosing the unitary gauge, $\theta(x) =0$, one can neglect the phase
of the waterfall in the background and perturbations analysis. One can also assume without loss of generality that the background gauge field points along the $x$ direction, with background components $A_{\mu}=(0,A(t),0,0)$. The system then reduces to a Bianchi type I universe which is given by
\ba
\label{metric}
ds^2 &=& - dt^2 + e^{2\alpha(t)}\left( e^{-4\sigma(t)}d x^2 +e^{2\sigma(t)}(d y^2 +d z^2) \right)  \nonumber\\
&\equiv&   - dt^2 + a(t)^2 dx^2 + b(t)^2 (dy^2 + dz^2)\,.
\ea
Here $\alpha(t)$ denotes the average number of e-foldings, $\dot \alpha$ refers to the isotropic Hubble expansion rate and $\dot \sigma(t)$
represents the anisotropic expansion rate. \\
The total energy density from gauge field plus inflaton field is given by
\ba
\label{energy}
{\cal E}=V(\phi,\chi) + e^{-2 N + 4 \sigma} \left( \frac{1}{2}f^2(\phi) \dot
A^2+\frac{\mathbf{e}^2\chi^2}{2} A^2 \right) \, ,
\ea
where the scalar fields kinetic energy has been neglected (slow-roll limit).
In addition, to simplify the notation, we defined $\alpha(t) =N(t)$ (number of e-foldings). \\

Let us consider the vacuum dominated regime where $\chi$ is very heavy during inflation and reaches its instantaneous minimum $\chi=0$ very quickly after the beginning of inflation. Then similarly to the standard Hybrid inflation model, as soon as the inflaton field reaches its critical value, $\phi=\phi_c \equiv \frac{M}{g}$, the waterfall field becomes tachyonic and thus rolls down quickly to its global minimum $|\chi |=\mu\equiv M/\sqrt{\lambda}, \phi=0$. At that point, inflation ends abruptly. In our current setup, the coupling of the gauge field to the waterfall modifies the surface of end of inflation. More specifically, the effective mass of waterfall in our model is
\ba
\label{chi-mass}
\frac{\partial^2 {\cal E}}{\partial \chi^2}\large|_{\chi=0}  = g^2 (\phi^2 - \phi_c^2) + \mathbf{e}^2 e^{- 2 N+4\sigma} A^2 \, .
\ea
In the absence of the gauge field, the moment  of waterfall instability is determined  when $\phi= \phi_c = M/g$. However, in the presence of gauge field the condition of waterfall instability is modified. More specifically, the condition of waterfall phase transition from  Eq.~(\ref{chi-mass}) can be rewritten as
\ba
\label{transition}
\phi_f^2 + \frac{\mathbf{e}^2}{g^2} A_f^2  =\phi_c^2\,,
\ea
in which $\phi_f$ and $A_f$ refer to the field values at the surface of end of inflation.
In the following, we choose the convention in such a way that the time of end of inflation corresponds to $N\equiv N_f=0$ and we count the number of e-foldings in such a way that  $N(t)= -\int_t ^{t_f} dt H <0$.  In order to solve the flatness and the horizon problem in FRW cosmology, one needs at least 60 e-foldings so at the onset of inflation we need to have $N\equiv N_i \simeq -60$. \\

In the vacuum dominated regime so the potential driving inflation is approximately given by
\ba
\label{V-app}
V\simeq \frac{M^4}{4\lambda}+\frac{1}{2}m^2\phi^2  \, .
\ea
Next we consider the background evolution of inflaton and gauge fields (we refer the reader to \cite{Abolhasani:2013bpa} for more details on its derivation). Since the aim is using the $\delta N$ formalism, instead of presenting the evolution of the fields in terms of $N$, we present the number of e-folds in terms of either the $\phi$ or the $A_{\mu}$ field. The final result is 
\ba
\label{N-phi-hybrid}
N(\phi) &\simeq& \frac{p_c}{2 \left(I- 1  \right) } \ln \left( \frac{\phi}{\phi_{f}} \right)  \\ \label{N-A-hybrid}
N(A) &\simeq& \frac{1}{3} \ln  \left( \sqrt{\frac{3}{2 R} } \left( \frac{ A -  A_{f} }{M_{P}} \right) + 1 \right) \, ,
\ea
where $p_c \equiv \left(\frac{M^4}{2 \lambda m^2 M_P^2} \right) $ and $R$ denotes the ratio of the energy density of the gauge field over the total energy density during inflation. $\phi_f$ and $A_f$ denote the values of the fields at the end of inflation. \\
A crucial assumption to get the above formulas is to consider the following ansatz for the gauge kinetic coupling, $f \simeq (a/a_f)^{-2} = e^{-2N}$. \\

The last step to obtain $\delta N$ is to account the contributions from the surface of the end of inflation in $\delta N$. We parameterize the surface of end of inflation in Eq.~(\ref{transition}) via \cite {Emami:2011yi}
\ba
\label{gamma}
\phi_{f} = \phi_c \cos \gamma \quad , \quad
A_{f} =  \frac{g\,  \phi_c}{\mathbf{e}} \sin \gamma \, .
\ea
Notice that $\gamma$ is an independent variable so in the computation of $\delta N$ one also needs to take into account the perturbations in $\gamma(\phi, A)$, i.e. the contribution of the surface of end of inflation in $\delta N$. \\
Having presented the background evolution of the fields as well as derived the surface of the end of inflation, we are ready to use the $\delta N$ formalism as well as cosmic perturbations. \\

\subsection{Power spectrum}
\label{power-spec}
\begin{figure}[!h]
	\centering
	\includegraphics[width=0.5\textwidth]{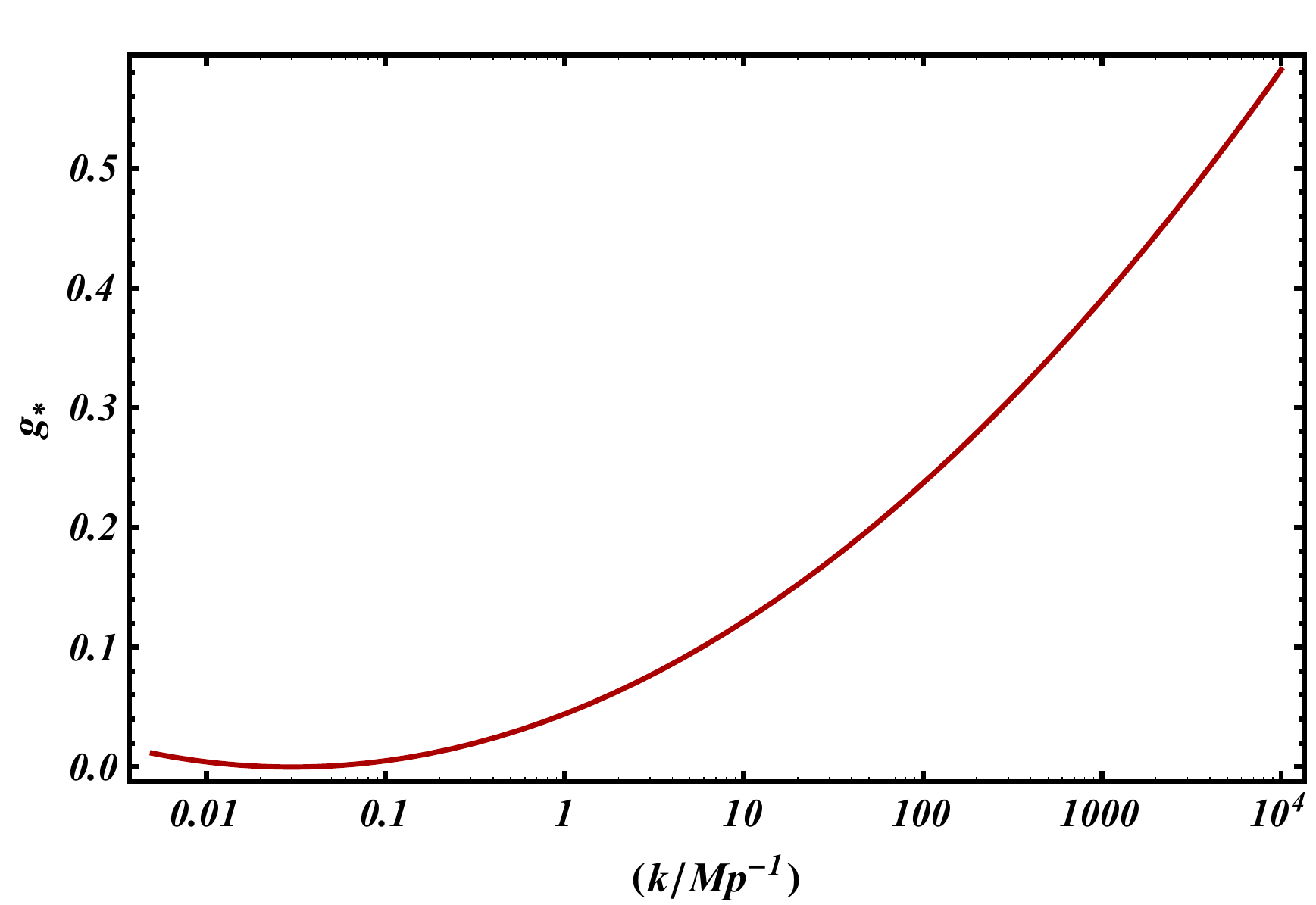}
	\caption{Evolution of $g_{*}$ as a function of scale. Here we assumed that  $(R/\epsilon) = 7.5 \times 10^{-5}$}
\label{gstar}
\end{figure}

As already mentioned, using the $\delta N$ formalism the curvature perturbation becomes $\R{({\mathbf{x}}, t)} = \delta N( \phi, A) $. Then it is straightforward to calculate the power spectrum of the curvature perturbation as
\ba
\P_{\R }= N_\phi^2 \P_{\delta \phi} + \dot A^2 N_{\dot A}^2 \P_{\delta \dot A/\dot A}
+ N_{A}^2 \P_{\delta A}\,,
\ea
where the expressions for the $ N_\phi$, $N_{\dot A}$ and $N_{A}$ are given in Appendix \ref{appB}. Using these expressions, we can rewrite the curvature power spectrum as
\ba
\P_{\R } &=& \P_{\R }^{(0)} \left(1 + \frac{\Delta \P_{\R}}{\P_{\R }^{(0)}}\right)\,, \\
\P_{\R }^{(0)} &=&  N_{\phi_*}^2 \P_{\delta \phi_*} = \left( \frac{p_c H_*}{4 \pi \phi_*}
\right)^2 \left[ 1+ \frac{\mathbf{e} p_c M_P}{g \phi_c} \frac{\tan \gamma}{\cos \gamma} \sqrt{\frac{2 R}{3}} \right]^2 \,,\\
\frac{\Delta \P_{\R}}{\P_{\R }^{(0)}} &=&
\left( \frac{\sqrt{\frac{48 R}{\epsilon}} N{(k)} - \frac{\mathbf{e}}{g} \tan \gamma}{1+ \frac{\mathbf{e} p_c M_P}{g \phi_c} \frac{\tan \gamma}{\cos \gamma} \sqrt{\frac{2}{3} R}} \right)^2 \sin^2 \theta \nonumber\\
&\simeq& \left( \sqrt{\frac{48 R}{\epsilon}} N{(k)} - \frac{\mathbf{e}}{g} \tan \gamma \right)^2  \sin^2 \theta\,,
\ea
where the slow-roll parameter, $\epsilon$, is given by
\ba
\label{epsilon-eq}
\epsilon \simeq \frac{2 \phi_f^2}{p_c^2 M_P^2}\,.
\ea
Notice that the corrected power spectrum has two contributions, one form the gauge kinetic coupling, the other from the surface of the end of inflation. It is then possible to choose the parameters in such a way as to produce a cancellation between these two contributions at some specific scales, which here we select it to be within the CMB scale, from $ 0.005 Mp^{-1} \leq  k \leq  0.1  Mp^{-1}$ (in the following we take for instance $ k = 0.03 Mpc^{-1}$). In order to cancel out the above mentioned contributions we would need
\ba
\label{cancellation}
\sqrt{\frac{48 R}{\epsilon}} N{(k = 0.03 Mpc^{-1})} = \frac{\mathbf{e}}{g} \tan \gamma
\ea
Furthermore, we assume that $ N{(k = 0.03 Mpc^{-1})} =-60$. For the next references we define $N^{*} \equiv N{(k = 0.03 Mpc^{-1})}$. \\
At any scale we can rewrite the corrected power spectrum as
\ba
\label{corrected power}
g_{*} &\equiv& -\frac{\Delta \P_{\R}}{\P_{\R }^{(0)}} \left(\frac{1}{\sin^2{\theta}}\right) = \left(\frac{48 R}{\epsilon} \right)\left(N(k) - N^{*} \right)^2\,.
\ea
Requiring $g_*< 0.01$ within the CMB window, one finds an upper bound $\frac{ R}{\epsilon} < 10^{-4}$. \\ 
Fig.~\ref{gstar} show a plot of $g_{*}$ as a function of $k$ in the whole window of the CMB and of the smaller scales. 
%%%%%%%%%%%%%%%%%%%%%%%%%%%%%%%%%%%%%%%%%%%%%%%%%%%%%
%%%%%%%%%%%%%%%%%%%%%%%%%%%%%%%%%%%%%%%%%%%%%%%%%%%%%

\subsection{Bispectrum}

\begin{figure}[!h]
	\centering
	\includegraphics[width=0.5\textwidth]{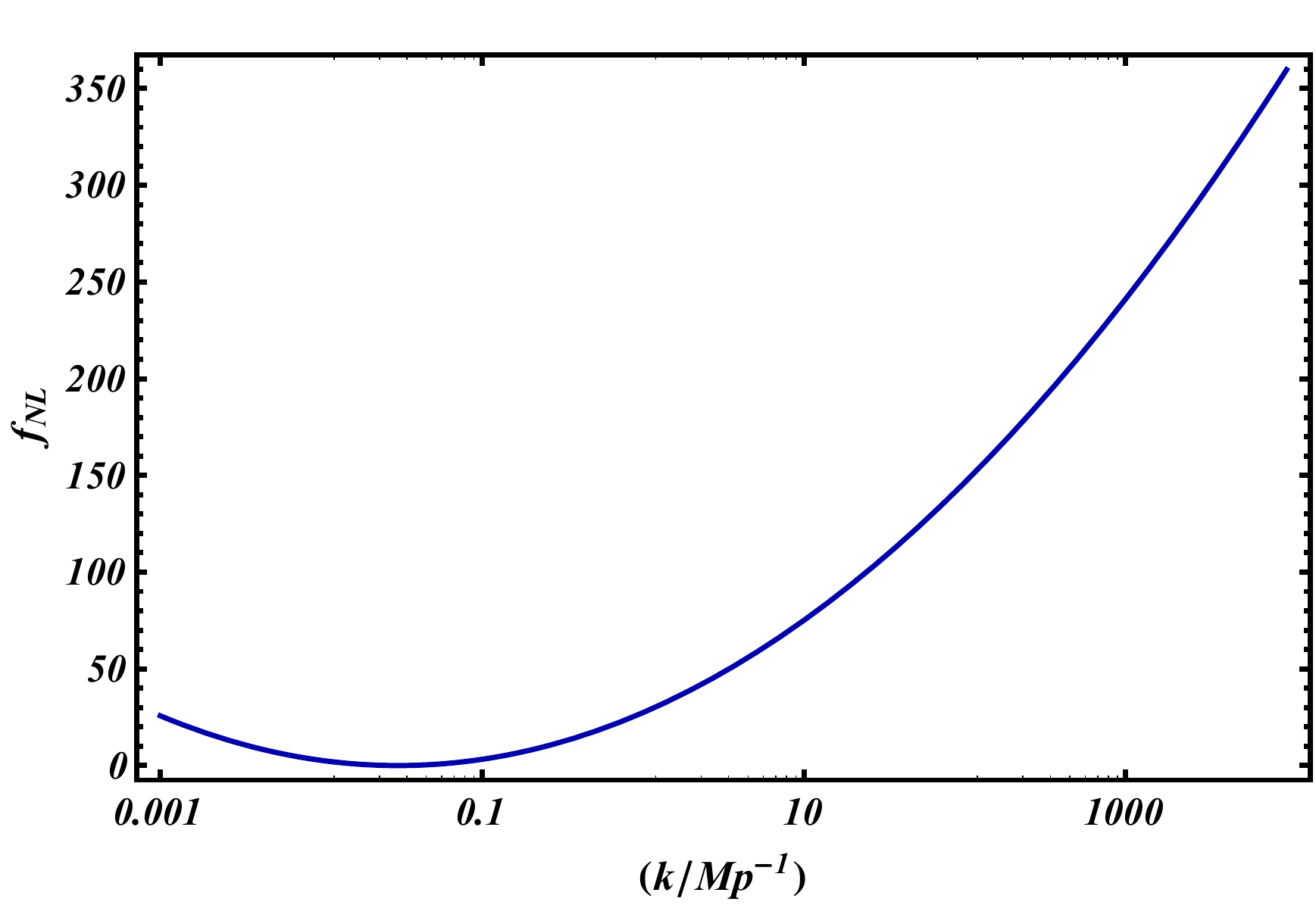}
	\caption{\label{fNlgauge}Evolution of $f_{\text{nl}}$ in the squeezed limit as a function of scale. We fix $k_3 = 0.005 \text{Mpc}^{-1}$.} 
	\label{FIG6}
\end{figure}

The result for the bispectrum is \cite{Abolhasani:2013bpa} 
\ba
\label{leading three}
B_{\R}(\mathbf{k}_1, \mathbf{k}_2, \mathbf{k}_3)   &\simeq& \left(\frac{3 \epsilon}{R}\right) \Big[ \sqrt{g_*(k_1) g_*(k_2)}
 \Big(\frac{2 R}{\epsilon}N_{k_3} - \frac{\mathbf{e}^2 \,  R \, p_{c}}{6 g^2}\frac{(\sin^2{\gamma}+1)}{\cos^4{\gamma}} \frac{M_{P}^2}{\phi_{c}^2}   \nonumber\\&-& \frac{ \mathbf{e}\, 2 R p_{c}}{3 g \epsilon}\frac{\tan{\gamma}}{\cos{\gamma}} \frac{M_{P}}{\phi_{c}}\sqrt{6R} \Big)\Big( C( \mathbf{k}_1, \mathbf{k}_2)P_0(k_{1})P_0(k_{2}) \Big)+ 2 \mathrm{perm.} \Big] \,  \nonumber\\
&\simeq& 6 \Big( \sqrt{g_*(k_1) g_*(k_2)} N_{k_3} \Big) \Big( C( \mathbf{k}_1, \mathbf{k}_2)P_0(k_{1})P_0(k_{2}) \Big)+ 2 \mathrm{perm.} \Big)\,,
\ea
where $C( \mathbf{k}_1, \mathbf{k}_2)$ is given by
\ba
C(\mathbf{k}_1, \mathbf{k}_2)\equiv
\bigg{(}1 -  (\widehat{\mathbf{k}_1}.\widehat {\mathbf{n}} )^2 
 - (\widehat{\mathbf{k}_2}.\widehat{\mathbf{n}})^2 +
 (\widehat{\mathbf{k}_1}.\widehat{\mathbf{n}}) \,  (\widehat{\mathbf{k}_2}.\widehat{\mathbf{n}}) \,  (\widehat{\mathbf{k}_1}.\widehat{\mathbf{k}_2} )  
 \bigg{)} \, .
\ea
Using the standard definition of $f_{\text{nl}}$ 
\ba
f_{\text{nl}} (\mathbf{k}_1, \mathbf{k}_2, \mathbf{k}_3) = \lim_{k_1 \rightarrow 0} \frac{5}{12}
\frac{B_\zeta(\mathbf{k}_1, \mathbf{k}_2, \mathbf{k}_3)}{P_\zeta(k_1) P_\zeta(k_2)} \, ,
\ea
one obtains its value in the squeezed limit
\ba
\label{squeezed limit}
f_{\text{nl}}\Big{|}_{k_3 \ll k_1 \simeq k_2} &=& - 10 g_{*}(k_1) N(k_3) C( \mathbf{k}_1, \mathbf{k}_2) \simeq  - 10 g_{*}(k_1) N(k_3) \,,
\ea
where we have neglected an $\mathcal{O}(1)$ term coming from $C( \mathbf{k}_1, \mathbf{k}_2)$. Due to the numerical behavior of $g_*$, which goes up toward the small scales, we have selected $k_1$ to be associated with the short mode, while the number of e-folds to be associated with the long mode. In order to study the CMB spectral distortion, we should select the short mode within the $\mu$ or $y$ scale and the long mode in the CMB temperature anisotropies window (here we select $k_3 = 0.005 Mpc^{-1}$). \\
The numerical behavior of $f_{\text{nl}}$ is presented in Fig.~\ref{fNlgauge}: an enhancement on $f_{\text{nl}}$ is achieved towards the smaller scales.\\
%%%%%%%%%%%%%%%%%%%%%%%%%%%%%%%%%%%%%%%%%%%%%%%%%%%%%%%%%%%%%%%%%%%%
%%%%%%%%%%%%%%%%%%%%%%%%%%%%%%%%%%%%%%%%%%%%%%%%%%%%%%%%%%%%%%%%%%%%

\section{Conclusions}
\label{conclusions}

Probing cosmological correlations in the squeezed limit is crucial for gaining insight into the microphysics of inflation, as well as testing its alternatives, and for understanding how to correctly map observations to models. \\
\indent CMB spectral distortions provide the  opportunity to test  inflation on a range of scales that is complementary w.r.t. the ones we have access to thanks to CMB temperature and polarization anisotropies or through the LSS. Primordial fluctuations re-entering Hubble during the $\mu$ or y-distortion era undergo dissipation by photon diffusion, resulting in a distortion of the CMB black-body spectrum. Correlating distortion and temperature anisotropies is a promising novel direction for testing primordial non-Gaussianity in the squeezed limit. Remarkably, this is done on scales smaller than those to which the current bounds on $f_{\text{nl}}$ apply. \\
\indent Inflationary models predicting (i) a correlation in the squeezed limit for cosmological perturbations and (ii) a larger value for $f_{\text{nl}}$ in the squeezed limit on $\mu$ or y-distortion scales than the current Planck bounds place on large scales, may be efficiently constrained with an experiment having detection limits at the level of the proposed PIXIE.\\
\indent  We discussed a variety of mechanisms that satisfy the two conditions above and derived quantitative predictions for the squeezed bispectrum on small scales for two concrete inflationary realizations that are representative of large classes of models. One of these is a scenario where a correlation among long and short-wavelength modes is generated via a resonance between sub-horizon inflaton fluctuations and an oscillatory background. The latter is specifically due to heavy fields (coupled to the inflaton) that oscillate and decay, generating bumps in the correlation functions at specific scales. We find that the squeezed non-Gaussian signal in this model is above detection limits on y-distortion scales (between $1$ and $50\,\text{Mpc}^{-1}$), without violating the Planck bounds. The signal is weaker on $\mu$ scales: non-Gaussianity is enhanced in so-called ``not-so squeezed" configurations, resulting in a $y-T$ correlation that can be much larger than $\mu-T$. Another example for which we evaluate the order of magnitude of the squeezed bispectrum is the \textit{gauged hybrid} model, another multi-field scenario, but one in which the production of a squeezed correlation is determined by a completely different mechanism: a time-dependent kinetic mixing between the inflaton and another light field, within a hybrid inflation set-up. In this case $f_{\text{nl}}$ grows monotonically at smaller scales, resulting in a signal that is largest on $\mu$ scales. \\

\indent As we commented at length in Sec.~(\ref{general2}), there exist other interesting and well-motivated scenarios that would be ideal candidates for experimental tests by means of distortion-temperature correlations. It would be interesting to perform for these models a quantitative analysis similar to the one that has been presented for the resonance and for the gauged hybrid models. One should also point out that, similarly to what happens on large scales, also on small scales information on  $f_{\text{nl}}$ would be very helpful in disentangling inflationary models that are degenerate in their predictions for the two point function but show qualitative differences at the level of higher order statistics.

\section*{Acknowledgments}
We are indebted to Marc Kamionkowski for asking the questions that initiated this project and for his collaboration at the early stages of this work. Thanks a lot also to Jens Chluba for the insightful discussions. We are delighted to thank the Cosmology group at JHU for very warm hospitality while this work was initiated. RE was supported by the New College Oxford-Johns Hopkins Centre for Cosmological Studies during her visit at JHU. This work was supported at ASU by the Department of Energy and at Hong Kong University by the CRF Grants of the Government of the Hong Kong SAR under HKUST4/CRF/13G. RE is also very grateful to IPMU and YITP Institutes for warm hospitality when this work was in its final stage.

\appendix

\section{Derivation of correlation functions in the resonance model}
\label{details}

\noindent The expectation value of an operator $\Theta$ at time $t$ can be
computed using the in-in formula \cite{inin}
\ba\label{ininf}
\left\langle \Theta(t) \right\rangle=\left\langle 0\,\Big| \left[\bar{T}\,\exp\left(i\int_{t_{0}}^{t}dt^{'}H_{I}(t^{'})\right)\right] \Theta_{I}(t)\left[T\,\exp\left(-i\int_{t_{0}}^{t}dt^{''}H_{I}(t^{''})\right)\right] \Big|\,0 \right\rangle\,.
\ea
This will be employed for the power spectrum and for the bispectrum computations in the resonance model, by expanding the interaction Hamiltonian respectively to quadratic and cubic order.\\
\indent The amplitude of the leading-order power spectrum contribution from the derivative interactions is proportional to the following integral 
\begin{equation}\label{ttt}
A(k)\sim \int_{0}^{t} dt' \Omega(t')\sin(2k\tau')\,,
\end{equation}
where $\Omega(t')$ indicates the background value of the massive field $\chi$ (or its time derivative, $\dot{\chi}$, depending on which of the $K_{i}$ interactions one considers), oscillating with frequency $m$, and the $\sin$ function in (\ref{ttt}) comes from the two mode functions in (\ref{cccc}), for which the subhorizon approximation has been employed to simplify the integrand. The position of the peak in the power spectrum and the corresponding wave number can be derived analytically. \\
\indent First one  computes the stationary point of the phase function in (\ref{ttt}) finding that the resonance occurs at time $t_{*}$ defined by $k/a(t_{*})=m/2$. The duration of the resonance is easily derived from the second derivative of the phase function, giving $\delta t\sim 1/\sqrt{Hm}$. This is a characteristic time-scale for the system, independent on the specific mode number. Integrating in the vicinity of the resonance, one obtains an analytic form for $A(k)$ that shows a characteristic peak at $k\approx k_{p}= m a_{0} e^{\sqrt{H/m}}$ (having set the scale factor equal to one at the onset of the massive field oscillation). Modes close to $k_{p}$ contribute with a larger correction to the power spectrum. \\
\indent Notice that this characteristic scale can be interpreted as that of the mode that resonates with the external frequency at time $t_{\text{system}}\simeq 1/\sqrt{Hm}$: the resonance condition is indeed $k/a(t_{*})=m$, from which it follows $k=k_{p}$ if $a(t_{*})\simeq e^{t_{\text{system}}}$. Notice also that consistency requires that $\Gamma<\sqrt{Hm}$ for the scale $k_{p}$ to lie in the interval of modes that can resonate with the external frequency. The smallest mode number to be affected by the resonance is equal to m at the onset of the oscillations, i.e. $k/a(t=0)=k_{\text{min}}=m$; the largest one reaches value m at the time when the massive field decays, $k_{\text{max}}/a(t_{\text{end}})=k/e^{H\Gamma^{-1}}=m$. One can see that $k_{p}>k_{\text{min}}$ always applies and $k_{p}<k_{\text{max}}$ is satisfied as long as the decay rate is sufficiently small, $\Gamma<\sqrt{mH}$.    \\

 The calculation for the bispectrum can be performed with the same techniques. The amplitude is proportional to an integral as in Eq.~(\ref{ttt}), where now $k\equiv k_{1}+k_{2}+k_{3}$. One finds that the amplitude has its maximum amplitude when $k\simeq k_{p}$. Looking at squeezed triangle configurations with $k_{1}\simeq k_{2}\ll k_{3}$, $k\simeq k_{p}$ implies $k_{1},k_{2}\simeq k_{p}/2$ and $k_{3}\ll k_{p}$. For modes to be subhorizon one needs to impose the condition $t_{i}>t_{*}$, where $t_{i}$ is the horizon crossing time for $k_{i}$, having set $i=1,2,3$, and $t_{*}$ is defined by $k/a(t_{*})\simeq m$ (this is the region from which the majority of the contribution for integrals as (\ref{ttt}) arises). The condition $t_{i}>t_{*}$ implies $a(t_{i})>a(t_{*})$, which in turn translates into $k_{i}/H>k/m\simeq k_{p}/m$ and therefore one can probe a resonance effect in the finitely squeezed (as opposed to arbitrarily squeezed) limit, i.e. for $k_{3}>k_{p} (H/m)$. \\

\noindent The third order interacting Hamiltonian from the $K_{p}$ term is given by
\begin{equation}
H^{(3)}_{I}=\int d^3 x \,a^3\,\left(\frac{\lambda_{p}\dot{\chi}}{4\Lambda_{j}^{4}}\right)\left[-\delta\dot{\phi}^3+\delta\dot{\phi}\frac{\left(\partial_{i}\delta\phi\right)^2}{a^2}\right]\,.
\end{equation}
The bispectrum to leading order is given by
\begin{eqnarray}\label{intapp}
\langle \delta\phi_{\vec{k}_{1}}\delta\phi_{\vec{k}_{2}}\delta\phi_{\vec{k}_{3}}\rangle_{H^{(3)}}&\simeq &(2\pi)^3 \delta^{(3)}(\vec{k}_{1}+\vec{k}_{2}+\vec{k}_{3})u_{k_{1}}(\tau)u_{k_{2}}(\tau)u_{k_{3}}(\tau)\times\nonumber\\&\times&\Big[3\,\alpha \,\int d\tau' a(\tau')\dot{\chi}\,u_{k_{1}}^{' *}(\tau')\,u_{k_{2}}^{' *}(\tau')\,u_{k_{3}}^{' *}(\tau')\nonumber\\&+&2\,\alpha \int d\tau'\,a(\tau')\dot{\chi}\,\vec{k}_{2}\cdot\vec{k}_{3}\,u_{k_{1}}^{' *}(\tau')\,u_{k_{2}}^{' *}(\tau')\,u_{k_{3}}^{' *}(\tau')\nonumber\\&+&\left[k_{1}\leftrightarrow k_{2}\right]+\left[k_{1}\leftrightarrow k_{3}\right]\Big]\,,
\end{eqnarray}
where $\alpha\equiv -\lambda_{p}/4\Lambda_{j}^{4}$ and the mode functions are given by
\begin{equation}
u_{k}(\tau)=\frac{i\,H}{\sqrt{2k^3}}\left(1+i k \tau\right)e^{-ik\tau}\,.
\end{equation}
The massive field background oscillates as $\chi(t)\simeq \chi_{0}e^{-\Gamma t}\cos(mt)$. Using the subhorizon approximation and approximating the full integrals in (\ref{intapp}) with their contributions from around the resonance, one finds
\begin{eqnarray} \label{comb1}
\langle \delta\phi_{\vec{k}_{1}}\delta\phi_{\vec{k}_{2}}\delta\phi_{\vec{k}_{3}}\rangle_{H^{(3)}}&\approx & \frac{m}{\Gamma}\cos(\theta_{*})\frac{\chi_{0}}{a_{*}^{3}}\frac{H^3 \lambda_{p}}{16\,\Lambda_{j}^{4}}\,F(k)\frac{1}{k_1 k_2 k_3}\nonumber\\ &\times& \left[3-\frac{k_1 \left(\vec{k}_{2}\cdot\vec{k}_{3}\right)+k_2 \left(\vec{k}_{1}\cdot\vec{k}_{3}\right)+k_3 \left(\vec{k}_{1}\cdot\vec{k}_{2}\right)}{k_1 k_2 k_3}\right]\,,
\end{eqnarray}
where $\theta_{*}\equiv (m/H)(1+\ln(k/a_{0}m))$ is the phase at the stationary point, $a_{*}\equiv a(t_{*})$ and the function $F$ is defined as
\begin{equation}
    F(k)\equiv
    \begin{cases}
      \left(\frac{k}{a_{0}m}\right)^{-\Gamma/m}\,e^{-\Gamma/\sqrt{mH}}-1, & \text{if}\ \, -\sqrt{\frac{m}{H}}<\frac{m}{H}\ln\left(\frac{k}{a_{0}m}\right)<\sqrt{\frac{m}{H}} \\
      \left(\frac{k}{a_{0}m}\right)^{-\Gamma/m}\,\left(e^{-\Gamma/\sqrt{mH}}-e^{\Gamma/\sqrt{mH}}\right), & \text{if}\ \,\frac{m}{H}\ln\left(\frac{k}{a_{0}m}\right)>\sqrt{\frac{m}{H}} \,.
    \end{cases}
  \end{equation}
The peak for the bispectrum occurs at $k\simeq k_{p}\equiv (a_0 m/2)e^{\sqrt{H/m}}$,  around which the function is well-approximated by $F(k)\approx e^{-2\Gamma/\sqrt{mH}}-1$, hence Eq.~(\ref{rrr}). We adopt the usual convention for the definition of the bispectrum amplitude
\begin{equation}\label{comb2}
\langle \zeta_{\vec{k}_{1}} \zeta_{\vec{k}_{2}}\zeta_{\vec{k}_{3}}  \rangle(2\pi)^{7}\delta^{(3)}(\vec{k}_{1}+\vec{k}_{2}+\vec{k}_{3})\frac{3}{10}f_{\text{nl}}\left(\mathcal{P}_{\zeta}\right)^{2}\frac{\sum_{i=1,2,3}k_{i}^{3}}{\Pi_{j=1,2,3}k_{j}^3}\,,
\end{equation}
where $\mathcal{P}_{\zeta}\equiv H^2/(8\pi^{2}M_{p}^{2}\epsilon_{V})$. Combining (\ref{comb1}) and (\ref{comb2}) and taking the limit of a finitely squeezed triangle ($k_{1}\simeq k_{2}\approx k_{p}\gg k_{3}>(H/m)k_{p}$), one arrives at Eq.~(\ref{comb3}).\\

Let us sketch a quick comparison among the contributions to the bispectrum from $K_{n}$, $K_{d}$  and $K_{p}$. From \cite{Saito} one has from $K_{n}$ (the same applies to $K_{d}$)
\begin{eqnarray}\label{deltabn}
\Delta \mathcal{B}^{(n)}&\approx &  \epsilon\frac{m}{H}\frac{\Delta\mathcal{P}_{\zeta}^{(n)}}{\mathcal{P}_{\zeta}}  \,,
\end{eqnarray}
where $\Delta\mathcal{P}^{(n)}_{\zeta}$ is the correction to the power spectrum due to $K_{n}$ itself and $\Delta \mathcal{B}$ is defined as a dimensionless bispectrum 
\begin{eqnarray}
\langle \zeta_{\vec{k}_{1}}\zeta_{\vec{k}_{2}}\zeta_{\vec{k}_{3}} \rangle\sim \frac{\mathcal{P}}{k_{1}^{2}k_{2}^{2}k_{2}^{3}}\Delta\mathcal{B}(k_{1},k_{2},k_{3})\,.
\end{eqnarray}
Using this definition along with our result in (\ref{comb3}), from $K_{p}$ we find
\begin{equation}\label{deltabp}
\Delta\mathcal{B}^{(p)}\approx \frac{k_{3}}{k_{p}}\left(\frac{m}{H}\right)^2  \frac{\Delta\mathcal{P}_{\zeta}^{(p)}}{\mathcal{P}_{\zeta}} \,,
\end{equation}
in the limit $k_p\simeq k_1\simeq k_2 \gg k_3$. Considering, for simplicity, the limiting case $\Delta\mathcal{P}_{\zeta}/\mathcal{P}_{\zeta}\approx 1$ both for (\ref{deltabn}) and (\ref{deltabp}), and using $k_3 >k_p (H/m)$, one finds $\Delta \mathcal{B}^{(p)}\gtrsim \mathcal{O}(1)\times(m/H)$, which is larger than (\ref{deltabn}).

\section{More details on Gauged hybrid inflation}
\label{appB}
Here we present the final expression for $ N_\phi$, $N_{\dot A}$ and $N_{A}$.
\begin{align}
\label{N-phi}
N_{{\phi}} & =   \frac{p_{c}}{2(I-1)} \frac{1}{{\phi}}   + \frac{\e M_{P}p_{c}}{ 6 \phi_c \, g \, (I-1)}\frac{\tan{\gamma}}{\cos{\gamma}} \frac{f_{\phi}}{f} \sqrt{ 6 R} \\
%%%%%%%%%%%%%%%%%%%%%%%%%%%%%%%%%%%%%%%%%%%%%%%%%
\label{N-dotA}
N_{\dot {A}} & = \left(-\frac{2NI}{(I-1)} +  \frac{ \e \, p_{c}}{6 g (I-1)}\frac{\tan{\gamma}}{\cos{\gamma}} \frac{M_{P}}{\phi_{c}} \sqrt{ 6 R} \right) \left( \frac{1}{\dot {A}}\right) \\
%%%%%%%%%%%%%%%%%%%%%%%%%%%%%%%%%%%%%%%%%%%%%%%%%
\label{N-A}
N_{{A}} & = \frac{\e \, p_{c}}{2 g \phi_c (I-1)}\frac{\tan{\gamma}}{\cos{\gamma}} \, .
\end{align}

%%%%%%%%%%%%%%%%%%%%%%%%%%%%%%%%%%%%%%%%%%%%%%%%%%%%%%%%%%%%%%%%%%%%
{}

\end{document}